\documentclass{iopart}

\usepackage{graphicx,epsfig,sidecap}
\usepackage{bm}

\def\be{\begin{equation}}
\def\ee{\end{equation}}

\def\bea{\begin{eqnarray}}
\def\eea{\end{eqnarray}}

\def\Tr{{\rm Tr}}

\def\e{\epsilon}

\def\HH{\mathcal H}
\def\vp{\varphi}

\newcommand{\rx}{{ x}}
\newcommand{\ry}{{\tau}}
\def\lt#1{\left#1}
\def\rt#1{\right#1}
\def\t#1{\tilde{#1}}

\newcommand{\C}{{\mathbf{C}}}

\newcommand{\bra}{\langle}
\newcommand{\ket}{\rangle}
\newcommand{\tw}{{\cal T}}

\begin{document}

\title[Entanglement entropy  of two disjoint intervals in CFT]
{Entanglement entropy of two disjoint intervals in conformal field theory}

\author{Pasquale Calabrese$^1$, John Cardy$^{2}$, and Erik Tonni$^1$}
\address{$^1$Dipartimento di Fisica dell'Universit\`a di Pisa and INFN,
             Pisa, Italy.\\
         $^2$ Oxford University, Rudolf Peierls Centre for
          Theoretical Physics, 1 Keble Road, Oxford, OX1 3NP, United Kingdom
          and All Souls College, Oxford.}

\date{\today}

\begin{abstract}

We study the entanglement of two disjoint intervals in the conformal field 
theory of the Luttinger liquid (free compactified boson).
$\Tr\rho_A^n$ for any integer $n$ is calculated as the four-point function 
of a particular type of twist fields and the final result is expressed in a 
compact form in terms of the Riemann-Siegel theta functions. 
In the decompactification limit we provide the analytic continuation valid 
for all model parameters and from this we extract the entanglement entropy. 
These predictions are checked against existing numerical data.

\end{abstract}

\maketitle

\section{Introduction}

The interest in quantifying the entanglement in extended quantum
systems has been growing in recent times at an impressive rate,
mainly because of its ability in detecting the scaling behaviour in
proximity of quantum critical points (see e.g. 
Refs. \cite{af-rev,e-rev,cc-rev,ccd-sp}  as reviews). 
A particularly useful measure of entanglement in the ground-state
of an extended quantum system is the entanglement entropy $S_A$. 
It is defined as follows. 
Let $\rho$ be the density matrix of a system, which we take to be in the 
pure quantum state $|\Psi\ket$, $\rho=|\Psi\rangle\langle\Psi|$.
Let the Hilbert space be written as a direct product $\HH=\HH_A\otimes\HH_B$. 
$A$'s reduced density matrix is $\rho_A=\Tr_B\, \rho$.
The entanglement entropy is the corresponding Von Neumann entropy
\be
S_A=-\Tr\, \rho_A \log \rho_A\,,
\label{Sdef}
\ee
and analogously for $S_B$. 
When $\rho$ corresponds to a pure quantum state $S_A=S_B$. 

The entanglement entropy revealed to be an optimal indicator of the 
critical properties of an extended quantum system when $A$ and $B$ correspond
to a spatial bipartition of the systems. 
Well-known and fundamental examples are critical one-dimensional systems in the 
case when $A$ is an interval of length $\ell$ in an infinite line.
In this case, the entanglement entropy follows the 
scaling \cite{Holzhey,Vidal,cc-04}
\be
S_A= \frac{c}3 \log\frac{\ell}a +c_1'\,,
\label{SAlog}
\ee
where $c$ is the central charge of the associated
conformal field theory (CFT) and $c'_1$ a non-universal constant.
Away from the critical point, $S_A$ saturates to a constant value
\cite{Vidal} proportional to the logarithm of the correlation length
\cite{cc-04}.
This scaling allows to locate the position (where $S_A$ diverges by 
increasing $\ell$) and a main feature (by the value of the central charge $c$) 
of quantum critical points displaying conformal invariance.

The central charge is an ubiquitous and fundamental characteristic of the 
CFT, but it does not always identify unequivocally the universality class
of the transition.
A relevant class of relativistic quantum field theories are the so-called  
Luttinger liquids, which describe an enormous number of 
physical systems of experimental and theoretical interest. 
Just to quote a few, the one-dimensional Bose gases with repulsive 
interaction, the (anisotropic) Heisenberg spin chains, carbon nanotubes
are all described by Luttinger liquid theory in their gapless phases.
Via bosonization, all these models can be written as free bosonic field 
theories with $c=1$. The different universality classes are 
distinguished by the compactification radius $R$ of the bosonic field, that 
corresponds to experimentally measurable critical exponents.

The entanglement entropy of a single block of length $\ell$, according 
to Eq. (\ref{SAlog}) is transparent to the value of the compactification 
radius, because it depends only on $c$. 
In Ref. \cite{fps-09} it has been shown that instead the entanglement entropy 
of disjoint intervals depends explicitely on $R$, and so it encodes universal 
properties of the CFT that are hidden in the entanglement of a single block.
(Oppositely in 2D systems with conformal invariant wave-function, the entanglement  entropy of a single region depends on $R$ \cite{2d}.)
Eq. (\ref{SAlog}) in a CFT is calculated by a modification of the replica 
trick of disordered systems \cite{Holzhey,cc-04}. In fact, one first calculates
$\Tr\rho_A^n$ for integral $n$, that results to be 
\be
\Tr\rho_A^n=c_n\left(\frac{\ell}a\right)^{-\frac{c}6 (n-1/n)}\,,
\label{Trn}
\ee
that is easily analytically continued to any complex value of $n$, 
and then $S_A=-\lim_{n\to 1} \partial_n \Tr\rho_A^n$ gives Eq. (\ref{SAlog}).
The reason of this way of proceeding is that for integral $n$,
$\Tr\rho_A^n$ is the partition function on an $n$-sheeted Riemann surface 
obtained by joining consecutively the $n$ sheets along region $A$ (see next 
section for details). We will refer to this surface as ${\cal R}_{n,N}$, 
where $N$ is the number of disjoint intervals composing $A$.
(${\cal R}_{n,N}$ are fully defined by the $2N$ branch points $u_j$ and $v_j$). 
In the case of a single interval, ${\cal R}_{n,1}$ is easily uniformised to the 
complex plane by a simple conformal mapping. Then the powerful tools of 
CFT give Eq. (\ref{Trn}).

However, when the subsystem $A$ consists of several disjoint intervals, the 
analysis becomes more complicated. 
In Ref. \cite{cc-04} (and also in \cite{cc-05p}), 
based on a uniformising transformation mapping 
${\cal R}_{n,N}$ into the complex plane, a general result for $\Tr\rho_A^n$
has been given. However, this result is in general incorrect. 
In fact, the surface ${\cal R}_{n,N}$ has genus $(n-1)(N-1)$ and so for 
$N\neq1$ cannot be uniformised to the complex plane (at the level of the 
transformation itself, this has been discussed in some details \cite{cg-08}). 
The case $n=N=2$ has the topology of a torus, whose partition function depends 
on the whole operator content of the theory and not only on the central 
charge (see e.g. \cite{cftbook}).
Consequently the simple formulas of Ref. \cite{cc-04} cannot be generally 
correct. The partition functions on Riemann surfaces with higher genus are
even more complicated.

We consider here the case of two disjoint intervals 
$A=A_1\cup A_2=[u_1,v_1]\cup [u_2,v_2]$
defining the surface ${\cal R}_{n,2}$. By global conformal invariance 
$\Tr \rho_A^n$ can be written as
\be\fl
\Tr \rho_A^n\equiv Z_{\mathcal{R}_{n,2}}
=c_n^2 \left(\frac{|u_1-u_2||v_1-v_2|}{|u_1-v_1||u_2-v_2||u_1-v_2||u_2-v_1|} 
\right)^{\frac{c}6(n-1/n)} {\cal F}_{n}(x)\,,
\label{Fn}
\ee 
where $x$ is the four-point ratio (for real $u_j$ and $v_j$, $x$ is real) 
\begin{equation}
x=\frac{(u_1-v_1)(u_2-v_2)}{(u_1-u_2)(v_1-v_2)}\,.
\label{4pR}
\end{equation}
This can be written as 
\be
Z_{\mathcal{R}_{n,2}}= Z_{\mathcal{R}_{n,2}}^W {\cal F}_{n}(x)\,,
\ee
where $Z_{\mathcal{R}_{n,2}}^W$ is the incorrect result in Ref. \cite{cc-04}.
We normalised such that ${\cal F}_{n}(0)=1$.
The function ${\cal F}_{n}(x)$ depends explictly on the full operator content 
of the theory and must be calculated case by case.

In Ref. \cite{fps-09}, using old results of CFT on orbifolded space 
\cite{Dixon,z-87}, ${\cal F}_{2}(x)$ has been calculated for a free boson 
compactified on a circle of radius $R$
\be
{\cal F}_{2}(x)=
\frac{\theta_3 (\eta \tau) \theta_3 (\tau/\eta)}{ [\theta_3 (\tau)]^{2}},
\label{F2}
\ee
where $\tau$ is pure-imaginary, and is related to $x$ via
$x= [\theta_2(\tau)/\theta_3(\tau)]^4$. 
$\theta_\nu$ are Jacobi theta functions.
$\eta$ is a universal critical exponent proportional to the square of the 
compactification radius $R$ (in Luttinger liquid literature $\eta=1/(2K)$).

The main result of this paper is ${\cal F}_{n}(x)$ for generic 
integral $n\geq1$:
\begin{equation}
\mathcal{F}_n(x)=
\frac{\Theta\big(0|\eta\Gamma\big)\,\Theta\big(0|\Gamma/\eta\big)}{
[\Theta\big(0|\Gamma\big)]^2}\,,
\label{Fnv}
\end{equation}
where $\Gamma$ is an $(n-1)\times(n-1)$ matrix with elements
\be
\Gamma_{rs} =  
\frac{2i}{n} \sum_{k\,=\,1}^{n-1} 
\sin\left(\pi\frac{k}{n}\right)\beta_{k/n}\cos\left[2\pi\frac{k}{n}(r-s)
\right]\,, 
\label{Gammadef}
\ee
and 
\be
\beta_y=\frac{F_{y}(1-x)}{F_{y}(x)}\,,\qquad
F_{y}(x)\,\equiv\, _2 F_1(y,1-y;1;x)\,.
\label{betadef}
\ee
$\eta$ is exactly the same as above, while $\Theta$ is the Riemann-Siegel 
theta function
\begin{equation}
\label{theta Riemann def}
\Theta(z|\Gamma)\,\equiv\,
\sum_{m \,\in\,\mathbf{Z}^{n-1}}
\exp\big[\,i\pi\,m^{\rm t}\cdot \Gamma \cdot m+2\pi i m^{\rm t}\cdot z\,\big]\,,
\end{equation}
with $z$ a generic complex vector with $n-1$ components.
$\Theta(0|\Gamma)$ for $n-1=1$ reduces to the Jacobi 
$\theta_3(\tau=i\beta_{1/2})$, 
and so Eq. (\ref{Fnv}) reproduces Eq. (\ref{F2}).
In the above, and hereafter, 
the dot ($\cdot$) denotes the matrix product and the superscript t 
($ ^{\rm t}$) the transposition.
Eq. (\ref{Fnv}) is manifestly invariant under $\eta\to 1/\eta$, as numerically
observed \cite{fps-09}. It is also invariant under $x\to 1-x$ (even if not 
manifest in this form). 
For $\eta=1$ we have ${\cal F}_n(x)=1$ and the result $Z^W_{{\cal R}_{2,N}}$ 
in \cite{cc-04} is then correct. This equality carries over to the 
analytic continuation and then to the entanglement entropy, confirming
what observed numerically \cite{fps-09}.

Unfortunately we have been not yet able to analytically continue this 
result to real $n$ for general values of $\eta$ and $x$, and so to obtain the 
entanglement entropy. However we managed to give asymptotic expressions 
for small and large $\eta$ that compare well with numerics. 

It is worth to recall that in the case of two intervals, the 
entanglement entropy measures only the entanglement of the 
two intervals with the rest of the system. 
It is {\it not} a measure of the entanglement of one interval with respect 
to the other, that instead requires the introduction of more complicated 
quantities because $A_1\cup A_2$ is in a mixed state (see e.g. Refs. 
\cite{Neg,Neg2} for a discussion of this and examples). 

The paper is organized as follows. 
In Sec. \ref{secRS} we recall how to obtain the entanglement entropy 
within CFT on Riemann surfaces and the usefulness of twist fields. 
In Sec. \ref{secCF} we introduce the free compactified boson and fix all 
our notation. 
The following section \ref{sec4P} is the core of paper where Eq. (\ref{Fn})
is derived. This section requires a good knowledge of CFT on orbifolds 
(to make the paper self-contained we also have a long \ref{apporb}
on this, where the results of Ref. \cite{Dixon} that we used are explained).
The reader uninterested in the derivation can skip this section, to read 
directly Sec. \ref{secAn} where a partial analytic continuation of Eq. 
(\ref{Fn}) is performed and the consequences are discussed with particular 
attention to the comparison with the numerical results in Ref. \cite{fps-09}.
Several appendices contain most of the technical parts of the paper.

\section{Entanglement entropy and Riemann surfaces}
\label{secRS}

Given a quantum field theory whose dynamics is described by the Hamiltonian 
$H$, the density matrix $\rho$ in a thermal state at inverse 
temperature $\beta$ may be written as a path integral in the imaginary time 
interval $(0,\beta)$
\begin{equation}
\label{pathi}
\fl \rho(\{\phi_x\}|\{\phi'_{x'}\})=
Z^{-1}\int[d\phi(y,\tau)]
\prod_x\delta(\phi(y,0)-\phi'_{ x'})\prod_x
\delta(\phi(y,\beta)-\phi_x)\,e^{-S_E}\,,
\end{equation}
where $Z(\beta)={\rm Tr}\,e^{-\beta H}$ is the partition function,
the euclidean action is $S_E=\int_0^\beta  L d\tau$, 
with $L$ the euclidean lagrangian. 
Here the rows and columns of the density matrix are labelled by the 
values of the fields at $\tau=0,\beta$.

The normalisation factor of the partition function ensures that
${\rm Tr}\rho=1$, and is found by 
setting $\{\phi_x\}=\{\phi'_{x}\}$ and integrating over these variables. 
In the path integral, this has the effect of sewing together the edges 
along $\tau=0$ and $\tau=\beta$ to form a cylinder of circumference $\beta$. 
Now let $A$ be a subsystem consisting of the points $x$ in the
disjoint intervals $(u_1,v_1),\ldots,(u_N,v_N)$. An expression for the
the reduced density matrix $\rho_A$ is obtained from (\ref{pathi})
by sewing together only those points $x$ which are not in $A$. This
has the effect of leaving open cuts, one for each interval
$(u_j,v_j)$, along the line $\tau=0$.

\begin{figure}
\includegraphics[width=\textwidth]{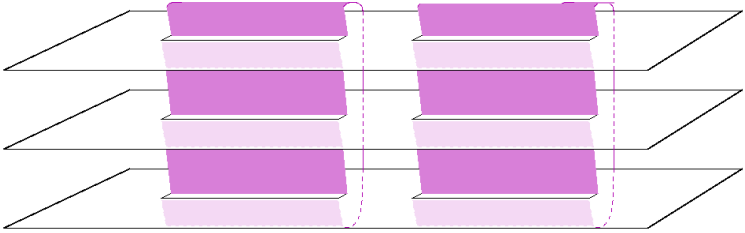}
\caption{A representation of the Riemann surface ${\cal R}_{3,2}$.} 
\label{fig-sheets}
\end{figure}

We may then compute ${\rm Tr}\,\rho_A^n$, for any positive
integer $n$, by making $n$ copies of the above, labelled by an integer
$j$ with $1\leq j\leq n$, and sewing them together cyclically along the 
the cuts so that $\phi_j(x,\tau=\beta^-)=\phi_{j+1}(x,\tau=0^+)$ and
$\phi_n(x,\tau=\beta^-)=\phi_1(x,\tau=0^+)$ for all $x\in A$. 
This defines an $n$-sheeted Riemann surface depicted for $n=3$ 
and in the case when $A$ is formed by two disjoint intervals 
in Fig. \ref{fig-sheets}.
The partition function on this surface will be  denoted by $Z_n(A)$ and so
\begin{equation}
\label{ZoverZ}
{\rm Tr}\,\rho_A^n={Z_n(A)\over Z^n}\,.
\end{equation}
When the right hand side of the above equation has a unique
analytic continuation to ${\rm Re}\,n>1$, its first derivative
at $n=1$ gives the required entropy
\begin{equation}
S_A=-\lim_{n\to1}{\partial\over\partial n}{\rm Tr}\,\rho_A^n=
-\lim_{n\to1}{\partial\over\partial n}{Z_n(A)\over Z^n}\,.
\end{equation}
Notice that even before taking the replica limit, these partition functions 
give the R\'enyi entropies 
\be
S^{(n)}_A= \frac1{1-n} \log {\rm Tr}\,\rho_A^n\,.
\ee

Since the lagrangian density does not depend explicitly on the Riemann
surface ${\cal R}_{n,N}$ as a consequence of its locality, 
it is expected that the
partition function can be expressed as an object calculated from
a model on the complex plane $\C$, where the structure of the Riemann surface 
is implemented through appropriate boundary conditions around the
points with non-zero curvature.
Consider for instance the simple Riemann surface ${\cal R}_{n,1}$ needed 
for the calculation of 
the entanglement entropy of a single interval $[u_1,v_1]$, made of $n$ sheets
sequentially joined to each other on the segment $x\in[u_1,v_1],\,\tau=0$. 
We expect that the associated partition function in a theory 
defined on the complex plane $z=x+i\tau$
can be written in terms of certain ``fields'' at $z=v_1$ and $z=u_1$.

The partition function (here  ${\cal L}[\varphi](z,{\bar z})$ is the 
local lagrangian density)
\be
\label{partfunct}
Z_{{\cal R}_{n,N}}= \int [d\varphi]_{{\cal R}_{n,N}} 
\exp\lt[-\int_{{\cal R}_{n,N}} d z d{\bar z}\,{\cal L}[\varphi](z,{\bar z})\rt]\,,
\ee 
essentially defines these fields (i.e. it gives their
correlation functions, up to a normalisation independent of their
positions). In order to work with local fields (see for an extensive 
discussion \cite{ccd-07}), it is useful to move
the intricate topology of the world-sheet (i.e. the space where the coordinates
$x,\ry$ lie) ${\cal R}_{n,N}$ to the target space (i.e. the space where the 
fields lie).
Let us consider a model formed by $n$ independent copies of the original 
model.
The partition function  (\ref{partfunct}) can be re-written as
the path integral on the complex plane
\be
\label{partfunctmulti}
\fl Z_{{\cal R}_{n,N}} =
\int_{{\cal C}_{u_j,v_j}} \,[d\varphi_1 \cdots d\varphi_n]_{\bf C} 
\exp\lt[-\int_{\bf C} 
d z d {\bar z}\,({\cal L}[\varphi_1](z,{\bar z})+\ldots
+{\cal L}[\varphi_n](z,{\bar z}))\rt]\,,
\ee
where with $\int_{{\cal C}_{u_j,v_j}}$ we indicated the 
{\it restricted} the path integral with conditions
\be
\fl    
{\cal C}_{u_j,v_j}:\quad
\varphi_i(\rx,0^+) = \varphi_{i+1}(\rx,0^-)~,\quad 
\rx\in\bigcup_{j=1}^N [u_j,v_j],\quad i=1,\ldots,n\,,
\label{Cdef}
\ee 
where we identify $n+i\equiv i$. The lagrangian density of the
multi-copy model is
\[
{\cal L}^{(n)}[\varphi_1,\ldots,\varphi_n](\rx,\ry) = 
{\cal L}[\varphi_1](\rx,\ry)+\ldots+{\cal L}[\varphi_n](\rx,\ry),
\]
so that the energy density is the sum of the energy
densities of the $n$ individual copies. Hence the expression
(\ref{partfunctmulti}) does indeed define local fields at
$(u_1,0)$ and $(v_1,0)$ in the multi-copy model \cite{ccd-07}.

The local fields defined in (\ref{partfunctmulti}) are examples of 
``twist fields''. 
Twist fields exist in a QFT model whenever there is a global internal symmetry 
$\sigma$ (a symmetry that acts the same way everywhere in space, 
and that does not change the positions of fields): 
$\int d\rx d\ry\, {\cal L}[\sigma\varphi](\rx,\ry) = 
\int d\rx d\ry\, {\cal L}[\varphi](\rx,\ry)$. 
In the model with lagrangian ${\cal L}^{(n)}$, there is a symmetry under 
exchange of the copies. The twist fields defined by (\ref{partfunctmulti}), 
which have been called {\em branch-point twist fields} \cite{ccd-07}, are 
twist fields associated to the two opposite cyclic permutation symmetries 
$i\mapsto i+1$ and $i+1\mapsto i$  ($i=1,\ldots,n,\;n+1\equiv 1$). 
We can denote them simply by $\tw_n$ and $\t\tw_n$, respectively
\bea 
\tw_n\equiv\tw_\sigma~,\quad &\sigma&\;:\; i\mapsto i+1 \ {\rm mod} \,n\,, \\
\t\tw_n\equiv\tw_{\sigma^{-1}}~,\quad &\sigma^{-1}&\;:\; i+1\mapsto i \ {\rm mod} \,n\,.
\eea
Notice that $\t\tw_n$ can be identified with $\tw_{-n}$.
Thus for the $n$-sheeted Riemann surface along the set $A$ made of 
$N$ disjoint intervals $[u_j,v_j]$ we have
\be
\label{pftwopt}
Z_{{\cal R}_{n,N}} \propto \bra 
\tw_n(u_1,0) \t\tw_n(v_1,0)\cdots \tw_n(u_N,0) \t\tw_n(v_N,0) \ket_{{\cal L}^{(n)},\C}\,.
\ee
This can be seen by observing that for $\rx\in[u_j,v_j]$, consecutive copies 
are connected through $\ry=0$ due to the presence of $\tw_n(v_j,0)$, 
whereas for $\rx$ in $B$, copies are connected to themselves through 
$\ry=0$ because the conditions arising from the definition of 
$\tw_n(u_j,0)$ and $\t\tw_n(v_j,0)$ cancel each other.
More generally, the identification holds for correlation functions in the 
model ${\cal L}$ on ${\cal R}$
\be\label{br-tw}
\fl \bra \Or(\rx,\ry; \mbox{sheet $i$}) \ket_{{\cal L},{\cal R}_{n,N}} =
 \frac{\bra \tw_n(u_1,0) \t\tw_n(v_1,0) \cdots\tw_n(u_N,0) \t\tw_n(v_N,0) \Or_i(\rx,\ry) \ket_{{\cal L}^{(n)},\C}}{\bra \tw_n(u_1,0) \t\tw_n(v_1,0) \cdots \tw_n(u_N,0) \t\tw_n(v_N,0)\ket_{{\cal L}^{(n)},\C}},
\ee
where $\Or_i$ is the field in the model ${\cal L}^{(n)}$ coming from 
the $i^{\rm th}$ copy of ${\cal L}$, and the ratio properly takes into account
all the proportionality constants.

It is often useful to introduce the linear combinations of the basic fields
\begin{equation}
\tilde{\varphi}_k \,\equiv\, 
\sum_{j\,=\,1}^n e^{2\pi i \frac{k}{n} j} \varphi_{j}\,,\qquad
k\,=\,0,1, \dots , n-1\,,
\end{equation}
that get multiplied by $e^{2\pi ik/n}$ on going around the twist operator, 
i.e. they diagonalize the twist
\be
\tw_n \tilde{\varphi}_k=e^{2\pi ik/n} \tilde{\varphi}_k\,,\qquad
\t\tw_n \tilde{\varphi}_k=e^{-2\pi ik/n} \tilde{\varphi}_k\,.
\ee
Notice that when the basic fields $\varphi_j$ are real then 
$\tilde{\varphi}_k^* = \tilde{\varphi}_{n-k}$. When the different 
values of $k$ decouple, the total partition function is a product of the 
partition functions for each $k$.
Thus also the twist fields can be written as products of fields acting 
only on  $\tilde{\varphi}_k$
\be
\tw_n=\prod_{k=0}^{n-1}\tw_{n,k}\,,\qquad 
\t\tw_n=\prod_{k=0}^{n-1}\t\tw_{n,k}\,,
\ee
with $\tw_{k,n} \tilde{\varphi}_{k'}= \tilde{\varphi}_{k'}$ if $k\neq k'$ 
and $\tw_{k,n} \tilde{\varphi}_{k}= e^{2\pi ik/n} \tilde{\varphi}_k$.
In terms of these fields, the partition function on the initial $n$-sheeted 
surface is 
\be\fl
Z_{{\cal R}_{n,N}}=\prod_{k=0}^{n-1} \bra \tw_{k,n}(u_1,0)\t\tw_{k,n}(v_1,0)
\dots \tw_n(u_N,0) \t\tw_n(v_N,0) \ket_{{\cal L}^{(n)},\C}\,.
\label{Tdiag}
\ee
This way of proceeding is very useful for free theories
when the various $k$-modes decouple leading to Eq. (\ref{Tdiag}). 
However, as we shall see soon, this is not straightforward for the problems
we are considering here, because the compactification condition introduces
a non trivial coupling between the $k$-modes.

\section{Compactified boson on a Riemann surface}
\label{secCF}

We consider a {\it complex} bosonic free field with Euclidean Lagrangian
density
\be
{\cal L}=\frac{g}{4\pi} |\nabla \vp|^2 \,.
\ee
The replicated theory is then ($\vp_j= \vp_{j,1} + i \vp_{j,2}$)
$$
{\cal L}^{(n)}= \frac{g}{4\pi} \sum_{j=1}^{n} 
\big( \partial_\mu  \vp_{j,1} \, \partial^\mu  \vp_{j,1}
+  \partial_\mu \vp_{j,2} \, \partial^\mu  \vp_{j,2}
\big)
= \frac{g}{2\pi n}  \sum_{k\,=\,0}^{n-1}  
\big( 
\partial_z  \tilde{\vp}_k  \, \partial_{\bar{z}}  \tilde{\vp}_k^\ast
+ \partial_z  \tilde{\vp}_k^\ast \, \partial_{\bar{z}}  \tilde{\vp}_k
\big)\,,
$$
with partition function
\begin{eqnarray}
\fl Z_{{\cal R}_{n,N}} &=&
\int_{\cal C} \left(\prod_{j\,=\,0}^{n-1} 
[d\vp_j]\right)
\,\exp\left\{\,-\frac{g}{4\pi} \sum_{j\,=\,0}^{n-1} \,\int  
\big( \partial_\mu  \vp_{j,1}  \partial^\mu  \vp_{j,1}
+  \partial_\mu \vp_{j,2}  \partial^\mu  \vp_{j,2}
\big) \,d^2 z\, \right\} \nonumber
\\ \fl
&=&
 \prod_{k=0}^{n-1}\, \int_{{\cal C}_k} 
[d\t\vp_k] [d\t \vp^*_k]
\exp\left\{\,-\frac{g}{2\pi\,n} \,\int  
\big( 
\partial_z  \tilde{\vp}_k  \, \partial_{\bar{z}}  \tilde{\vp}_k^\ast
+ \partial_z  \tilde{\vp}_k^\ast \, \partial_{\bar{z}}  \tilde{\vp}_k
\big) \,d^2 z\, \right\}\,,
\end{eqnarray}
where ${\cal C}$ stands for the restriction conditions in (\ref{Cdef})
and ${\cal C}_k$ are the corresponding conditions on the fields $\t\vp_k$.

The fields $\vp_j$ are free, but with each component compactified on a circle.
Since the field is complex, the target space of ${\cal L}$ is a torus with 
radii $R_1$ and $R_2$ .
Encircling the branch points can lead to a non-trivial winding that 
can be written as  (in the case of a branch point at the origin)
\begin{equation}\fl
\vp_j(e^{2\pi i}z, e^{-2\pi i}\bar{z})\,=\,
\vp_{j-1}(z, \bar{z})+   R_1 m_{j,1}+i R_2 m_{j,2}\,,
\qquad m_{j,1} , m_{j,2}\,\in\,\mathbf{Z}\,,
\end{equation}
where ${\bf Z}$ is the set of integer numbers. 
In the following we will only consider equal compactification radii 
$R\equiv R_1=R_2$.
For the fields $\tilde{\vp}_k$  these conditions read
\begin{equation}
\fl
\label{phik compactification}
\tilde{\vp}_k (e^{2\pi i}z, e^{-2\pi i}\bar{z})
=
e^{2\pi i \frac{k}{n}}  \tilde{\vp}_k (z, \bar{z})
+ R \sum_{j\,=\,1}^n e^{2\pi i \frac{k}{n} j} m_j
=
\theta_k \tilde{\vp}_k (z, \bar{z})+ R \sum_{j\,=\,1}^n \theta_k^j m_j\,,
\end{equation}
where $m_j \in \mathbf{Z}+i\mathbf{Z}$ and we introduced 
$\theta_k \equiv e^{2\pi i \frac{k}{n}}$.
Consequently, for a given $k\neq 0$, the target space for the fields $\t\vp_k$ 
is compactified on a complicated ``lattice''  $R  \Lambda_{{k}/{n}}$,
where $\Lambda_{{k}/{n}}$ is the two dimensional set of vectors given by 
the generic combination through integer coefficients of the vectors 
$\{ \theta_k,\theta_k^2, \dots, \theta_k^n \}$ (we recall that 
$\theta_k^n =1$), i.e.
\begin{equation}
 \Lambda_{\frac{k}{n}}=
\left\{
q\,=  \sum_{j\,=\,0}^{n-1} \theta_k^j m_j\,
\; ; \; m_j \in \mathbf{Z}+i \mathbf{Z}
 \right\}\,.
\label{lambdakn}
\end{equation}
This complicated structure of $\Lambda_{k/n}$ defining the space where 
the fields $\t\vp_k$ are compactified is the precise reason why the 
approach based on Eq. (\ref{Tdiag}) becomes difficult for the calculation of 
$Z_{{\cal R}_{n,2}}$. 
Few simple remarks on the structure of $\Lambda_{k/n}$ are in order. 
For fixed $n$, $\Lambda_{k/n}=\Lambda_{1-k/n}$.
The vectors $\theta_k^j$ are not independent (e.g. 
$\sum_{j=0}^{n-1}\theta_k^j=0$). 
%
In the following we will mainly use the results by Dixon et al. \cite{Dixon},
but we have to stress here that there are a series of papers from 
late eighties about conformal field theories on 
orbifold (e.g. \cite{z-87,k-87,br-87,otherorb})
that are very useful for the problem at hands and even 
for more complicated cases.  
The strategy of Ref. \cite{Dixon} to calculate these partition functions 
consists in splitting the field $\tilde{\vp}_k$ in a classical and a 
quantum part $\tilde{\vp}_k=\tilde{\vp}_k^{\rm cl}+\tilde{\vp}_k^{\rm qu}$.
The evaluation of the Gaussian functional integral naturally divides into a 
sum over all classical solutions, times exponential of 
the quantum effective action.
The classical solution takes into account the non-trivial topology of the 
target space as
\begin{equation}
\tilde{\vp}_k^{\rm cl}(e^{2\pi iz}z,e^{-2\pi i\bar{z}}\bar{z})= 
\theta_k^j \tilde{\vp}_k^{\rm cl}(z,\bar{z})+v\,,
\end{equation}
where $v\in R \Lambda_{k/n}$,
while the quantum fluctuations are transparent to it
\begin{equation}
\tilde{\vp}_k^{\rm qu}(e^{2\pi iz}z,e^{-2\pi i\bar{z}}\bar{z})= 
\theta_k^j \tilde{\vp}_k^{\rm qu}(z,\bar{z})\,.
\end{equation}

\subsection{The two-point function and entanglement of a single interval}

The two-point function of the twist fields 
for a complex field $\vp$ is \cite{Dixon} (see also \ref{apporb})
\begin{equation}
\langle \tw_{k,n}(u) \t\tw_{k,n}(v)  \rangle \,
\propto\,
\frac{1}{|u-v|^{4\Delta_{k/n}}}\,,
\end{equation}
where the dimensions of the twist fields read 
\begin{equation}
\Delta_{\frac{k}{n}} =\bar{\Delta}_{\frac{k}{n}}  =
\frac{1}{2}\,\frac{k}{n}\left(1-\frac{k}{n}\,\right)\,.
\end{equation}
Using Eq. (\ref{Tdiag}), the partition function on $\mathcal{R}_{n,1}$ is
\begin{equation}
Z_{{\cal R}_{n,1}}
\,=\,\prod_{k=0}^{n-1} Z_{k,n}
=\prod_{k=0}^{n-1}\langle \tw_{k,n}(u) \t\tw_{k,n}(v)  \rangle 
=\frac{c_n}{|u-v|^{4x_n}}\,,
\end{equation}
with
\begin{equation}
\label{x_n}
x_n\,=\,\sum_{k=0}^{n-1} \Delta_{\frac{k}{n}}  \,=\,
\frac{1}{12}\left(n-\frac{1}{n}\,\right)\,,
\end{equation}
and we defined the normalisation constant $c_n$ according to \cite{cc-04}.
This is in agreement with the direct calculation \cite{cc-04} for 
central charge $c=2$ because we are dealing with a complex field.
For a {\it real field} the correlation function is the square root
of the previous result, leading to an exponent $x_n$ that is the half of above,
in agreement with a $c=1$ theory.

In the case of the two-point function it is then very easy to use Eq. 
(\ref{Tdiag}) to obtain $Z_{{\cal R}_{n,1}}$. This is not the same for 
$Z_{{\cal R}_{n,2}}$ and we will adopt a slightly different strategy.

\section{The four-point function of twist fields}
\label{sec4P}

Using Eq. (\ref{Tdiag}) one is tempted to write the partition function on 
${\cal R}_{n,2}$ as
\begin{equation}\fl
\label{Z two intervals}
Z_{\mathcal{R}_{n,2}}
=\prod_{k=0}^{n-1} Z_{k,n}=\prod_{k=0}^{n-1}
\langle \tw_{k,n}(u_1,0) \t\tw_{k,n}(v_1,0) \tw_{k,n}(u_2,0) \t\tw_{k,n}(v_2,0)
\rangle\,.
\end{equation}
This choice is consistent with the requirement that there should be no 
monodromy on  going around both $u_1$ and $v_1$ or both $u_2$ and $v_2$.

Using global conformal invariance the four-point function of twist fields 
can be generally written as 
\begin{equation}
\fl Z_{k,n}(u_1,v_1,u_2,v_2)\propto
\left(\frac{|u_1-u_2||v_1-v_2|}{|u_1-v_1||u_2-v_2||u_1-v_2||u_2-v_1|} 
\right)^{4\Delta_{k/n}} {\cal G}_{k,n}(x)\,,
\label{Gkn}
\end{equation}
where $x$ is the four-point ratio given in Eq. (\ref{4pR}).
In the case of interest here $x$ is real, but in the most general case 
can take any complex value (see also \cite{cg-08}).
From Eq. (\ref{Gkn}) $Z_{\mathcal{R}_{n,2}}$ reduces to Eq. (\ref{Fn}),
after we normalise such that ${\cal F}_{n}(0)=1$ (for $x\to0$ 
$Z_{\mathcal{R}_{n,2}}$ is the product of two two-point functions
previously calculated and normalised with $c_n$).

However, because of the compactification conditions of the field 
$\t\vp_k$ Eq. (\ref{phik compactification}), it is not easy to write 
the classical contributions to $Z_{k,n}$ and to avoid multiple-counting of 
the various classical solutions when summing over them. 
For this reason, we will adopt a mixed strategy. 
We will use Eq. (\ref{Z two intervals}) for the quantum part of 
$Z_{{\cal R}_{n,2}}$ that is transparent to the compactification conditions. 
This will allow to re-use the results of Ref. \cite{Dixon} without 
modification. For the classical contribution instead, we will sum over all
the possible configurations of the fields $\vp_j$ that have easier 
compactification condition than $\t\vp_k$.

\subsection{The quantum part}

The quantum part of the four point correlation (\ref{Gkn})
is independent from the compactification of the target space and 
it is responsible for the full scaling factor. 
It has been calculated in Ref. \cite{Dixon} and we report its derivation in 
the \ref{apporb}. The final result is Eq. (\ref{Zqu 2int}) that for real 
$x$ becomes   
\begin{equation}\fl
Z_{k,n}^{\rm qu}= {\rm const}
\left(\frac{|u_1-u_2||v_1-v_2|}{|u_1-v_1||u_2-v_2||u_1-v_2||u_2-v_1|} 
\right)^{4\Delta_{k/n}}
\frac{1}{I_{k/n}(x)}\,,
\label{Zqu}
\end{equation}
where
\begin{equation}
I_{k/n}(x)\equiv
2F_{k/n}(x) F_{k/n}(1-x)
=2 \beta_{k/n}[F_{k/n}(x)]^2\,,
\label{I22}
\end{equation}
where $F_{y}(x)$ and $\beta_{k/n}(x)$ are given in Eq. (\ref{betadef}). 
Compared to Ref. \cite{Dixon} and to the appendix we have been stressing all 
the dependence on $k/n$ of the various functions.
Note that Eq. (\ref{I22}) is manifestly invariant for $x\to 1-x$.
From this expression, the contribution of the quantum fluctuations to 
$Z_{{\cal R}_{n,2}}$ is readily obtained from Eq. (\ref{Z two intervals}).

\subsection{The classical part}

The value of the action on the classical solution of the equation of motion 
contributing to $Z^{\rm cl}_{k}$ has been derived in 
Ref. \cite{Dixon} for a general orbifolded theory. 
This derivation is reported in full details in the \ref{apporb}. 
Here, for simplicity in the calculation and for physical reasons, 
we specialize to the case of $x$ real.
The action for a given classical configuration can be read from 
Eq. (\ref{dixon Scl}) in the appendix (fixing $\alpha_{k/n}=0$ and using our 
normalisation for the action):
\begin{equation}
S_{\rm cl}= \frac{2g \pi\sin{(\pi k/n)}}{n\beta_{k/n}} \big[|\xi_2|^2
+ \beta_{k/n}^2 |\xi_1|^2\big]\,,
\label{Scl}
\end{equation}
where 
$\xi_1, \xi_2 \in R\Lambda_{k/n}$ are generic vectors of the target space 
lattice $R\Lambda_{k/n}$. 
At this point we could calculate $Z_{k,n}$ by summing over the vectors
in the target space $R\Lambda_{k/n}$ given by Eq. (\ref{lambdakn}).
This computation would be very difficult. But not only, if in fact we would 
have been able to make this sum, the total partition function 
$Z_{\mathcal{R}_{n,2}}$ would not be given by Eq. (\ref{Z two intervals})
because some classical solutions would be counted more than once.

For all these reasons, we prefer to calculate the total partition function 
$Z_{\mathcal{R}_{n,2}}$ as the sum over all the classical configurations 
independently from the value of $k$.
Thus, for a complex field compactified in both directions with the same 
radius, we have 
\begin{equation}
Z_{\mathcal{R}_{n,2}} =
\sum_{m \in  \mathbf{Z}^{2n}} \prod_{k=0}^{n-1} Z^{\rm qu}_{k,n} Z^{\rm cl}_{k,n}\,, 
\end{equation}
and so in the total ${\cal F}_n(x)$ the scaling factor simplifies to give 
\be\fl
{\cal F}_n(x)=
\sum_{m \,\in \, \mathbf{Z}^{2n}}\prod_{k\,=\,0}^{n-1}\, 
\frac{\rm const}{\beta_{k/n}\big[F_{k/n}(x)\big]^2}
\,\exp\left\{-\frac{2g \pi\sin\left(\pi\frac{k}{n}\right)}{n} 
\left[|\xi_1|^2 \beta_{k/n} +\frac{|\xi_2|^2}{\beta_{k/n}}\right]
\right\}\,.
\ee
Notice that we keep $\sum_m$ out of $\prod_k$ because, for each $k$, all the 
components of the vector $m \in \mathbf{Z}^{2n}$ (we are dealing with the 
complex $\vp_j$'s; for real $\vp_j$'s $m \in \mathbf{Z}^{n}$) are involved 
in the condition (\ref{phik compactification}).
The quantum part does not depends on $m$, and so we can factor it out 
as anticipated above:
\begin{equation}\fl
\label{Z full}
\mathcal{F}_n(x) =
\left[\prod_{k=0}^{n-1} 
\frac{\rm const}{\beta_{\frac{k}{n}}[F_{\frac{k}{n}}(x)]^2}\right]
\sum_{m \in \mathbf{Z}^{2n}}
\prod_{k=0}^{n-1}\, 
\exp\left\{\frac{-2g \pi\sin\left(\pi\frac{k}{n}\right)}{n} 
\left[|\xi_1|^2 \beta_{\frac{k}{n}} +\frac{|\xi_2|^2}{\beta_{\frac{k}{n}}}\right]
\right\}\,.
\end{equation}
Given
\begin{equation}
\xi_p =
R \sum_{l=0}^{n-1}\theta_k^l ( m_{l,1}^{(p)}+i m_{l,2}^{(p)})\,,\qquad p=1,2,
\label{xip}
\end{equation}
we have
\begin{eqnarray}\fl
|\xi_p|^2&=&
R^2 \sum_{r,s=0}^{n-1} \left[\sum_{q=1,2}
m_{r,q}^{(p)} \cos\left[2\pi\frac{k}{n}(r-s)\right] m_{s,q}^{(p)} 
\right.\nonumber\\ \fl&&\left.\hspace{4cm}
+  (m_{r,1}^{(p)} m_{s,2}^{(p)}- m_{s,1}^{(p)} m_{r,2}^{(p)})
\sin\left[2\pi\frac{k}{n}(r-s)\right]
\right]\nonumber
\\ \fl
&\equiv&
R^2 \left[
\sum_{q=1,2}
\big[m^{(p)}_q\big]^{\rm t}  \cdot C_{\frac{k}{n}}\cdot m^{(p)}_q
+\sum_{r,s=0}^{n-1}(m_{r,1}^{(p)} m_{s,2}^{(p)}- m_{s,1}^{(p)} m_{r,2}^{(p)}) 
\big(S_{\frac{k}{n}}\big)_{rs}\right]\,,
\label{xicomp}
\end{eqnarray}
where $m^{(p)}_q \in \mathbf{Z}^n$ and 
\begin{equation}
\left( C_{\frac{k}{n}} \right)_{rs}\,\equiv\, 
 \cos\left[2\pi\frac{k}{n}(r-s)\right]\,,\quad
\left( S_{\frac{k}{n}} \right)_{rs}\,\equiv\, 
\sin\left[2\pi\frac{k}{n}(r-s)\right]\,.
\label{CCdef}
\end{equation}
Notice that $C_{\frac{k}{n}} $ is invariant for $k \leftrightarrow n-k$, 
while $S_{\frac{k}{n}}$ changes sign. 
For this reason, when summing over $k$ in the total partition function 
at fixed $r,s$, the two terms with  $S_{\frac{k}{n}}$ and  $S_{\frac{n-k}{n}}$
cancel out. This fundamental simplification does not happen if we 
consider only the partition sum at fixed $k$.
The remaining two sums over $m^{(p)}_q \in \mathbf{Z}^n$ factorize and we have
\begin{eqnarray}
\fl
\label{Z classical}
Z_{\rm cl} 
\label{riemann theta in Z}
&  =&
\left[\sum_{m \in  \mathbf{Z}^{n}}
\prod_{k\,=\,0}^{n-1}\, 
\exp\left\{-\frac{2\pi \,g}{n} \,R^2 \sin\left(\pi\frac{k}{n}\right)
\left[\, \beta_{k/n} \,m^{\rm t} \cdot C_{\frac{k}{n}}\cdot m +\frac{m^{\rm t} \cdot C_{\frac{k}{n}}\cdot m}{\beta_{k/n}}\,\right]
\, \right\}\right]^2\nonumber \\ \fl
\rule{0pt}{.8cm} &  =&\;
\left[\sum_{m \,\in \, \mathbf{Z}^{n}}
\exp\left\{\,i\,\pi
\Big[\, m^{\rm t} \cdot \Omega \cdot m + m^{\rm t} \cdot \widetilde{\Omega}\cdot m\,\Big]
\, \right\}\right]^2,
\end{eqnarray}
where the matrices $\Omega$ and $\widetilde{\Omega}$ are 
\begin{eqnarray}
\Omega_{rs} & \equiv & 
2 g R^2\;\frac{i}{n} \sum_{k\,=\,0}^{n-1} 
\sin\left(\pi\frac{k}{n}\right)  \beta_{\frac{k}{n}}
\cos\left[2\pi\frac{k}{n}(r-s)\right]\,,
\\ 
\widetilde{\Omega}_{rs} & \equiv & 
2 g R^2\;\frac{i}{n} \sum_{k\,=\,0}^{n-1} 
\sin\left(\pi\frac{k}{n}\right)  \frac{1}{\beta_{\frac{k}{n}}}
\cos\left[2\pi\frac{k}{n}(r-s)\right]\,,
\end{eqnarray}
and the indices $r$ and $s$ run over $1, \dots , n$. 
We remark that the term corresponding to $k=0$  is zero, thus we can deal 
with $\sum_{k=1}^{n-1}$ in these definitions. 
All elements of these matrices have vanishing real part.

Given a $G\times G$ symmetric complex matrix $\Gamma$ with positive 
imaginary part, the Riemann-Siegel theta function is defined as 
in Eq. (\ref{theta Riemann def}).
However, $\Omega$ and $\widetilde{\Omega}$ have one common eigenvector with vanishing eigenvalue that is $(1,1,\dots,1)$.
The sum in Eq. (\ref{Z classical}) is then divergent and cannot be written as
$\Theta (0| \Omega ) \Theta(0 | \widetilde{\Omega} )$. 
After diagonalizing these matrices, we will see
that is easy to adsorb this divergence in the normalisation factor.

The eigenvalues of $\Omega$ and $\widetilde{\Omega}$ are 
\be
\label{eigenvalue}
\fl
\omega_q 
=2 g R^2\sin\bigg(\pi\frac{q}{n}\bigg)  i\beta_{q/n},
\quad
\tilde{\omega}_q =2 g R^2\sin\bigg(\pi\frac{q}{n}\bigg)\frac{i}{\beta_{q/n}},
 \qquad q=1, \dots, n\,,
\ee
For $q=n$ (or equivalently $q=0$) the eigenvalue is vanishing, and then 
the imaginary parts of $\Omega$ and $\widetilde{\Omega}$ 
are not positive definite. 
Notice also that $\omega_q = \omega_{n-q}$ and $\t\omega_q = \t\omega_{n-q}$.

The matrices $\Omega$ and $\widetilde{\Omega}$ have a common eigenbasis whose normalised eigenvectors can be written as 
\begin{equation}
(y_q)_r\,\equiv\,\frac{e^{2\pi i \frac{q}{n} r}}{\sqrt{n}},
\hspace{1.5cm}
q , r\,=\,1, \dots , n.
\end{equation}
Thus, the $n \times n$ complex matrix $U$ whose 
elements are $U_{rs} \equiv (y_r)_s$ is unitary 
and it simultaneously diagonalizes $\Omega$ and $\widetilde{\Omega}$, i.e.
$$
U\Omega U^\dagger =
\left(\begin{array}{ccc|c}
\omega_1 & & & 0\\
 & \ddots & & \vdots \\
  &  & \omega_{n-1}& 0 \\
  \hline
   0 &  & 0& 0 
\end{array}
\right) \equiv\,\Omega_{\rm d},\quad
U \widetilde{\Omega} U^\dagger =
\left(\begin{array}{ccc|c}
\tilde{\omega}_1 & & & 0\\
 & \ddots & & \vdots \\
  &  & \tilde{\omega}_{n-1}& 0 \\
  \hline
   0 &  & 0& 0 
\end{array}
\right) \equiv\,\widetilde{\Omega}_{\rm d}.
$$
In order to extract the divergence from (\ref{Z classical}), we introduce a 
regulator by setting $i\e$ ($0<\e\ll 1$) instead of $0$ for the last 
eigenvalue i.e.
\begin{equation}\fl
\label{diagonal matrices eps}
\Omega_{{\rm d},\epsilon} \,\equiv\,
\left(\begin{array}{ccc|c}
\omega_1 & & & 0\\
 & \ddots & & \vdots \\
  &  & \omega_{n-1}& 0 \\
  \hline
   0 &  & 0& i \epsilon 
\end{array}
\right)\,,
\hspace{1.6cm}
\widetilde{\Omega}_{{\rm d},\epsilon} \,\equiv\,
\left(\begin{array}{ccc|c}
\tilde{\omega}_1 & & & 0\\
 & \ddots & & \vdots \\
  &  & \tilde{\omega}_{n-1}& 0 \\
  \hline
   0 &  & 0& i \epsilon 
\end{array}
\right) \,.
\end{equation}
We introduce 
$\widehat{U}$ as the restriction of $U$ to the first $n-1$ eigenvectors (i.e.  
we have dropped the eigenvector generating the kernel)
$\widehat{U}_{qr}\equiv {U}_{qr}$
for $q , r=1, \dots , n-1$.
We remark that $\widehat{U}$ is not unitary. 
We write the vector of integer numbers $m$ as $m=M+\hat m$, with $M$ belonging to 
the kernel (i.e. proportional to $(1,\dots,1)$) and $\hat m$ to the space orthogonal to it. 
Using the orthogonality of $M$ and $\hat m$, we have  
$$
m^{\rm t} \cdot \Omega \cdot m
= \big( \bar{U} m\big)^{\rm t} \cdot \Omega_{\rm d}\cdot\big( U m\big)
=[\lim_{\epsilon \,\rightarrow\,0}\,
( \bar{U} M)^{\rm t} \cdot \Omega_{\rm d,\e}\cdot( U M)]+
( \bar{U} \hat m)^{\rm t} \cdot \Omega_{\rm d}\cdot( U \hat m)
$$
where we explicitly used that in the space orthogonal to the kernel the product gives a finite 
result (and then the sum will be finite).
We have the same relation for $\widetilde\Omega$.
We can 
re-organize the sum in Eq. (\ref{riemann theta in Z}), summing before on the numbers 
spanned by $\hat m$ (that give a finite $\Theta$ function in an $n-1$ dimensional space) 
and after the diverging sum over the kernel. We finally find 
\begin{equation}
\label{Z classical theta3 eps}
Z_{\rm cl}\,=
\left[
\left(\, \lim_{\epsilon \,\rightarrow\,0} \frac1{ n\epsilon}\,\right)\,
\Theta\big(0|\eta\Gamma\big)\,\Theta\big(0|\eta\widetilde{\Gamma}\big)
\right]^2\,,
\end{equation}
where we defined
\begin{equation}\fl
\eta\Gamma \,\equiv\,
\widehat{U}^\dagger \left(\begin{array}{ccc}
\omega_1 & & \\
 & \ddots &  \\
  &  & \omega_{n-1} 
\end{array}
\right) \widehat{U}\,,
\hspace{1.5cm}
\eta\widetilde{\Gamma} \,\equiv\,
\widehat{U}^\dagger \left(\begin{array}{ccc}
\tilde{\omega}_1 & & \\
 & \ddots &  \\
  &  & \tilde{\omega}_{n-1} 
\end{array}
\right) \widehat{U}\,,
\end{equation}
which are symmetric and have positive imaginary parts and therefore provide 
well defined Riemann-Siegel theta functions.

The matrices $\eta\Gamma$ and $\eta\widetilde{\Gamma}$ are $\Omega$ and 
$\widetilde{\Omega}$ respectively with the last line and row dropped, i.e.
\begin{eqnarray}
\fl \Gamma_{rs} & = & 
\frac{2i}{n} \sum_{k\,=\,1}^{n-1} 
\sin\left(\pi\frac{k}{n}\right)\beta_{k/n}\cos\left[2\pi\frac{k}{n}(r-s)\right] 
=\frac{2}{n} \sum_{k\,=\,1}^{n-1}\sin\left(\pi\frac{k}{n}\right) i \beta_{k/n}\; e^{2\pi i \frac{k}{n}(r-s)} \,, \nonumber \\
\fl\widetilde{\Gamma}_{rs} & = & 
\frac{2i}{n} \sum_{k\,=\,1}^{n-1} 
\sin\left(\pi\frac{k}{n}\right) \frac{1}{\beta_{k/n}}\cos\left[2\pi\frac{k}{n}(r-s)\right] 
=\frac{2}{n} \sum_{k\,=\,1}^{n-1} \sin\left(\pi\frac{k}{n}\right) \frac{i}{\beta_{k/n}}\;e^{2\pi i \frac{k}{n}(r-s)}\,,\nonumber
\end{eqnarray}
where $r,s = 1, \dots, n-1$.
We introduced the matrices $\Gamma$ and $\widetilde{\Gamma}$, in such a way 
that they do not depend on $R$ and we defined 
\begin{equation}
\eta= g R^2 \,,
\end{equation}
that is exactly the same as in Ref. \cite{fps-09}, while the 
normalisation of $R$ is different.

Thus, taking (\ref{Z full}) and (\ref{Z classical theta3 eps}), adsorbing the divergence for $\epsilon \to 0$ 
into the constant, we have
\begin{equation}
\label{Z full 2}\fl
\mathcal{F}_n(x)=\,{\rm const}
\frac{[\Theta\big(0|\eta \Gamma\big)\,\Theta\big(0|\eta\widetilde{\Gamma}\big)]^2}{\prod_{k\,=\,1}^{n-1} \beta_{k/n}\big[F_{k/n}(x)\big]^2}=
\,{\rm const}
\frac{[\Theta\big(0|\eta \Gamma\big)\,\Theta\big(0|\eta\widetilde{\Gamma}\big)]^2}{\prod_{k\,=\,1}^{n-1} F_{k/n}(x)F_{k/n}(1-x) }
\,.
\end{equation}
This expression is manifestly invariant for $x\to 1-x$ because under this 
transformation $\beta_{k/n}\leftrightarrow 1/\beta_{k/n}$ 
and so $\Gamma\leftrightarrow \t\Gamma$.

In order to fix properly the normalisation constant (given by 
${\cal F}_n(0)=1$) and to show explicitly the invariance under $\eta\to1/\eta$,
it is worth to manipulate this expression.
Using Poisson resummation formula, in \ref{apppoi} we show
the following identity 
\begin{eqnarray}
\label{theta gamma tilde}
\Theta\big(0|\eta\widetilde{\Gamma}\big)&=&
\eta^{-\frac{n-1}{2}}\left(\,\prod_{k\,=\,0}^{n-1}\beta_{k/n}\right)^{\frac{1}{2}}
\,\Theta\big(0|\,\Gamma/\eta\big)\,.
\end{eqnarray}
We finally have 
\begin{equation}
\mathcal{F}_n(x)=
{\rm const}
\frac{[\Theta\big(0|\eta\Gamma\big)\,\Theta\big(0|\eta\widetilde{\Gamma}\big)]^2}{\prod_{k\,=\,1}^{n-1} \beta_{k/n}\big[F_{k/n}(x)\big]^2}=
\left[
\frac{\Theta\big(0|\eta\Gamma\big)\,\Theta\big(0|\Gamma/\eta\big)}{\prod_{k\,=\,1}^{n-1} F_{k/n}(x)}
\right]^2
\label{Fnp}
\end{equation}
where we fixed the constant by requiring that for ${\cal F}_n(0)=1$ 
( we used that for $x\to0$,  $F_{k/n}(0)=1$, and
$\beta_{k/n} \rightarrow + \infty$; furthermore 
$\Theta\big(0|\Gamma\big)$ goes to 1 for $\beta_{k/n}\to\infty$). 
We have thus written ${\cal F}_n$ in a way that is manifestly symmetric for 
the exchange $\eta \leftrightarrow1/\eta$.

A last manipulation can be done by using 
\begin{equation}
[\Theta(0|\Gamma)]^2=\prod_{k\,=\,1}^{n-1} F_{k/n}(x)\,,
\label{th1}
\end{equation}
proved in \ref{apptom}. 
This finally leads to the square of Eq. (\ref{Fnv}), in fact
this equation is valid for a complex field compactified in both 
directions with the same radius. The real field corresponds to the square root
of the previous result. This final manipulation allows to manifestly show 
that ${\cal F}_n(x)|_{\eta=1}=1$ identically.

It is also worth to mention that we wrote the partition function 
over a $n$-sheeted Riemann surface in terms of a Riemann-Siegel theta 
function defined with 
a matrix of dimension $n-1$ that is exactly the genus of the covering surface.

\subsection{Special cases}

\subsubsection{$n=2$.}

In this case the matrix $\Gamma$ is just 1 by 1, 
and so $\Theta(0|\Gamma) $ is a standard Jacobi $\theta_3$ function. 
Thus Eq. (\ref{Fnp}) ($\tau_{1/2}=i\beta_{1/2}$)
\begin{equation}
{\cal F}_{2}(x)=\left[
\frac{\theta_3(\tau_{1/2}\eta) \theta_3(\tau_{1/2}/\eta)}{ \theta_3^2(\tau_{1/2})}\right]^2\,,
\end{equation}
where we used that $F_{1/2}(x)=\theta_3^2(\tau_{1/2})$.
This is exactly the square of the result in Ref. \cite{fps-09} as it must be.

\subsubsection{$n=3$.}

First we observe that there is only one $\tau$ because 
$\tau_{1/3}= \tau_{2/3}$.  
We have
\begin{equation}\fl
\label{gamma prime n=3}
\Gamma\,=\,
\frac{\tau_{1/3}}{\sqrt{3}}
\left(\begin{array}{cc}
2 & -1 \\
-1 & 2
\end{array}\right)
\,=\,
\frac{1}{\sqrt{2}}
\left(\begin{array}{cc}
1 & -1 \\
1 & 1
\end{array}\right)
\left(\begin{array}{cc}
\gamma/3 & 0 \\
0 & \gamma
\end{array}\right)
\frac{1}{\sqrt{2}}
\left(\begin{array}{cc}
1 & -1 \\
1 & 1
\end{array}\right)^{\rm t}\,,
\end{equation}
where $\gamma = \sqrt{3}\,\tau_{1/3}$. Since these matrices are written in 
terms of integers only, this allows to write $\Theta\big(0| \eta\Gamma\big)$
as a sum of $\theta_3$ and $\theta_2$ and finally using the duplication 
formulas to prove the following identity
\begin{equation}\fl
\big[\Theta\big(0| \eta\Gamma\big)\big]^2
\,=\,\frac{1}{2}\Big[\,
\theta_2(\eta \gamma)^2  \theta_2\Big(\frac{\eta \gamma}3\Big)^2
+\theta_3(\eta \gamma)^2  \theta_3\Big(\frac{\eta \gamma}3\Big)^2
+\theta_4(\eta \gamma)^2  \theta_4\Big(\frac{\eta \gamma}3\Big)^2
\,\Big]\,.
\end{equation}
The same is clearly true for $\Theta\big(0| \Gamma/\eta\big)$, obtaining 
\begin{eqnarray}\fl
\label{full Z n=3}
{\cal F}_3(x) &=&
\frac{1}{4 [F_{1/3}(x)]^4}
\Big[\,
\theta_2(\eta \gamma)^2 \theta_2\Big(\frac{\eta \gamma}3\Big)^2
+\theta_3(\eta \gamma)^2 \theta_3\Big(\frac{\eta \gamma}3\Big)^2
+\theta_4(\eta \gamma)^2 \theta_4\Big(\frac{\eta \gamma}3\Big)^2
\,\Big]\nonumber\\ \fl
& &\times
\Big[\,
\theta_2\Big(\frac{\gamma}\eta\Big)^2 \theta_2\Big(\frac{\gamma}{3\eta}\Big)^2
+\theta_3\Big(\frac{\gamma}\eta\Big)^2  \theta_3\Big(\frac{\gamma}{3\eta}\Big)^2
+\theta_4\Big(\frac{\gamma}\eta\Big)^2  \theta_4\Big(\frac{\gamma}{3\eta}\Big)^2
\,\Big]\nonumber\,.
\end{eqnarray}

\subsubsection{$n=4$.}
Using tricks similar to the case $n=3$, it is possible to write 
$\Theta\big(0|\Gamma\big)^2$ for $n=4$ as 
a sum of 6 terms that are products of three $\theta_i$ (with different $\tau$'s
now, because in general $\tau_{2/4}\neq \tau_{1/4}=\tau_{3/4}$). 
This expression is however very long and not illuminating.

\subsection{Decompactification regime}

For fixed $x$, in the limit of large $\eta$ we have 
$\Theta\big(0|\eta\Gamma\big)=1 + \dots$ and
\begin{equation}
\Theta\big(0|\Gamma/\eta\big)\,=\,
\frac{1}{\sqrt{{\rm det}(-i\Gamma/\eta)}} \,\big(1+ \dots\big)
\,=\,
\frac{\eta^{(n-1)/2}}{\sqrt{{\rm det}(-i\Gamma)}} \,\big(1+ \dots\big)\,,
\end{equation}
where $\dots$ denotes vanishing terms as $\eta \rightarrow \infty$. 
Thus
\begin{equation}
{\cal F}_n(x)= \frac{\eta^{n-1}}{\prod_{k=1}^{n-1}F_{k/n}(x)F_{k/n}(1-x)}\,,
\label{decomp}
\end{equation}
recovering the correct quantum result Eq. (\ref{Zqu 2int}) 
with the proper  $\eta$ dependent normalisation (we used 
${\rm det}(-i\Gamma)=\prod_{k=1}^{n-1} \beta_{k/n}$ and
$\beta_{k/n}= F_{k/n}(1-x)/F_{k/n}(x)$).

Note that, using the symmetry $\eta\leftrightarrow 1/\eta$, the same formula 
gives also the answer for $\eta\to 0$.

\subsection{Small $x$ regime}
\label{Secsmallx}

Another case when the final result can be written in terms of simple functions
is the expansion for small $x$, corresponding to two far distant intervals 
(and by symmetry $x\to 1-x$ also for $x$ close to 1, corresponding to close 
intervals).  

For $x \to 0$ the expansion of $\beta_{k/n}$ is 
$$
 \beta_{\frac{k}{n}} =
-\frac{\sin\left(\pi\frac{k}{n}\right)}{\pi}
\left(\log x+ 
f_{\frac{k}{n}}
+\sum_{l=1}^{\infty} p_l\left(\frac{k}{n}\right)x^l\right),
\hspace{.2cm}
f_{\frac{k}{n}} \equiv 
2\gamma_E +\psi\left(\frac{k}{n}\right)+ \psi\left(1-\frac{k}{n}\right),
$$
where $\gamma_E$ is the Euler gamma, $\psi(z)\equiv\Gamma'(z)/\Gamma(z)$ is 
the Polygamma function (also know as digamma function) and $p_l(z)$ is a 
polynomial of degree $2l$, whose explicit expression is not needed. 
Plugging this expansion in the Riemann-Siegel theta function we obtain
\bea
\fl \Theta(0|\eta \Gamma) =1+\\  
\fl  \sum_{m \,\in\, {\bf Z}^{n-1} \setminus \{\vec{0}\}}
x^{\eta \frac{2}{n} \sum_{k=1}^{n-1}\sin\left(\pi\frac{k}{n}\right)^2\,
m^{\textrm{t}}\cdot \,C_{k/n} \cdot \,m}
e^{\eta \frac{2}{n} \sum_{k=1}^{n-1}\sin\left(\pi\frac{k}{n}\right)^2\,
f_{k/n} \, m^{\textrm{t}}\cdot \,C_{k/n} \cdot \,m}
\,\big(1+O(x)\big),\nonumber
\eea
where the matrix $C_{k/n}$ is defined in (\ref{CCdef}).
In this expansion the leading term is provided by those vectors 
$m \in  {\bf Z}^{n-1} \setminus \{\vec{0}\}$
which minimize the expression
\begin{equation}
\label{x exponent}
 \frac{2}{n} \sum_{k=1}^{n-1}\sin\left(\pi\frac{k}{n}\right)^2\,
m^{\textrm{t}}\cdot \,C_{k/n} \cdot \,m
\,=\, 
\sum_{j\,=\,1}^{n-1} m_j^2 -\sum_{j\,=\,1}^{n-2} m_j \,m_{j+1}\,.
\end{equation}
This expression is obviously minimized by all vectors of the form 
$m^{\textrm{t}}_{l,\pm} \equiv (0, \dots, 0, \pm1, \dots, \pm1, 0, \dots, 0)$,
with $l$ contiguous $\pm1$'s . 
At fixed $l=1, \dots , n-1$, there are $2(n-l)$ of such vectors 
for which the expression (\ref{x exponent}) is 1.
In order to evaluate the coefficient in front of the leading term, 
we observe that
\begin{equation}
m^{\textrm{t}}_{l,\pm} \cdot \,C_{k/n} \cdot \,m_{l,\pm} 
\,=\, \sum_{r,s\,=\,1}^{l} \left(C_{\frac{k}{n}} \right)_{rs}
\,=\,\left(\frac{\sin\left(\pi\frac{k}{n} l\right)}{\sin\left(\pi\frac{k}{n}\right)}\right)^2\,.
\end{equation}
Using the following integral representation for the digamma function
\begin{equation}
\psi(y) +\gamma_E= \int_0^\infty\frac{e^{-t}-e^{-yt}}{1-e^{-t}}\,dt\,,
\end{equation}
exchanging the order of sum and integral and then performing the latter, 
after simple algebra we find
\begin{equation}
\fl \frac{2}{n} \sum_{k\,=\,1}^{n-1}\sin\left(\pi\frac{k}{n}\right)^2\,
f_{k/n} \;  m^{\textrm{t}}_{l,\pm} \cdot \,C_{k/n} \cdot \,m_{l,\pm} 
\,=\,-\,\log\left[2n \sin\left(\pi\frac{l}{n}\right)\right]^2\,.
\end{equation}
Now, assuming $\eta \neq 1$, in the ratio of Riemann-Siegel theta functions occurring in 
$\mathcal{F}_n(x)$ the leading term is given by the minimum between $\eta$ and $1/\eta$, that 
we indicate as $\alpha={\rm min}(\eta,1/\eta))$ 
Therefore we get the small $x$ behavior of the scaling 
function
\begin{equation}
\fl\mathcal{F}_n(x) =1+
x^\alpha
\sum_{l=1}^{n-1}
\frac{2(n-l)}{\left[ 2n \sin\left(\pi\frac{l}{n}\right) \right]^{2\alpha}} + \dots=
1+ 2\left(\frac{x}{4n^2}\right)^\alpha
\sum_{l\,=\,1}^{n-1}
\frac{l}{\left[ \sin\left(\pi\frac{l}{n}\right) \right]^{2\alpha}}+ \dots
\,, \label{Fsmallx}
\end{equation}
where the dots denote higher order terms in $x$. Note that 
for $n=2$ this results  reduces to 
$\mathcal{F}_2(x) =1+2(x/16)^{{\rm min}(\eta,1/\eta)}+\dots$, 
as already found in \cite{fps-09}.
We stress, once more, that this expansion is valid only for $\eta\neq1$, in fact for 
$\eta=1$ the denominator in ${\cal F}_n(x)$ (that is of order $O(x)$) cancels exactly 
the numerator.

\subsection{Different compactification radii}

In this manuscript we mainly condisider the case of a complex boson 
compactified in both directions with the same radius, because it has
more physical applications and also to lighten the notation. 
However it is straightforward to generalize to the case with different 
compactification radii (at least for real four-point ratio $x$). 
The only change is that the target space is given by the product of 
two circles with different radii $R_1 \neq R_2$. 
Eq. (\ref{xip}) now becomes
\begin{equation}
\xi_p \,=\,
 \sum_{l=0}^{n-1}\theta_k^l \left( R_1 m_{l,1}^{(p)} + i \,R_2 m_{l,2}^{(p)}  \right)\,.
\end{equation}
The only (minimal) changes compared to the case of equal radii are 
in the computation in Eqs. (\ref{xicomp}). 
Repeating the straighiforward algebra we have
\begin{eqnarray}
\fl & & |\xi_p|^2 =
\sum_{r,s\,=\,0}^{n-1}\theta_k^{r} \bar{\theta}_k^s 
 \left( R_1m_{r,1}^{(p)} + i \,R_2m_{r,2}^{(p)}  \right)  \left( R_1m_{s,1}^{(p)} - i \,R_2m_{s,2}^{(p)}  \right) 
\\ \fl
& & =
\sum_{r,s\,=\,0}^{n-1}\left[\left( C_{\frac{k}{n}} \right)_{rs}
(R_1^2 m_{r,1}^{(p)} m_{s,1}^{(p)} +R_2^2  m_{r,2}^{(p)} m_{s,2}^{(p)})-
R_1 R_2\left( S_{\frac{k}{n}} \right)_{rs}
(m_{r,2}^{(p)} m_{s,1}^{(p)} - m_{r,1}^{(p)} m_{s,2}^{(p)}   )
\right]
\nonumber \\ \fl 
& &
=\;
\left[\,R_1^2\,m_1^{(p)\,\textrm{t}}  \cdot C_{\frac{k}{n}}\cdot m_1^{(p)}
+R_2^2\,m_2^{(p)\,\textrm{t}}  \cdot C_{\frac{k}{n}}\cdot m_2^{(p)}
+ R_1 R_2\, \textrm{tr}\left( A \,S_{\frac{k}{n}} \right)
\,\right]\,.
\nonumber
\end{eqnarray}
The important point is that the term in $R_1R_2$ is still vanishing and so 
the classical part of the action corresponds to the one of two independent real 
bosons. Note that this property is not completely trivial because the 
compactification conditions are not on the fields $\t\varphi_k$. 
The divergences due to $k=0$ are removed as before, arriving to 
the final expression
\be
\mathcal{F}_n(x)=\left[\frac{\Theta(0|\eta_1 \Gamma)\,\Theta(0| \Gamma/\eta_1)}{\Theta(0| \Gamma)^2}\right]
\left[\frac{\Theta(0|\eta_2 \Gamma)\,\Theta(0| \Gamma/\eta_2)}{\Theta(0| \Gamma)^2}\right]\,,
\ee
where $\eta_j \equiv g R^2_j$, the matrix $\Gamma$ is
defined in Eq. (\ref{Gammadef}) and the normalization has been chosen such
that $\mathcal{F}_n(0)=1$.

\section{The analytic continuation and the entanglement entropy}
\label{secAn}

In order to obtain the entanglement entropy we should be able to analytically 
continue Eq. (\ref{Fnv}) to general complex value of $n$ and, only after, take 
the derivative for $n\to1$. The definition of the Riemann-Siegel theta 
function makes this program hard, because it explictly involves summations
involving a matrix of dimensions $(n-1)\times (n-1)$ that are not obviously continued
to real values of $n$. One should find a different representation of the same 
function which is manipulable and we have been unable to do it. 
However, it is possible to analytically continue the denominator of 
Eq. (\ref{Fnv}). This is not only a first step towards the  full analytic 
continuation, but it provides all the answer in the decompactification regime (Eq.
\ref{decomp}), allowing to give precise predictions for $\eta\ll1$ and 
$\eta\gg1$. 

The logarithm of the denominator in Eq. (\ref{Fnp}) is a sum over 
$k=1\dots n-1$, which involves analytic functions of $k$. 
In this case it is possible 
to use the residue theorem to write the sum as an integral in the complex plane
of $\log F_{k/n}(x)$ times a function that has poles only for integer values 
of $k$ and over a contour that encircles all of them. 
A useful representation of this is 
\be
D_n(x)=\sum_{k=1}^{n-1}\log F_{k/n}(x)=
\int_{\cal C}\frac{dz}{2\pi i} \pi \cot(\pi z) \log F_{z/n}(x)\,,
\ee
where ${\cal C}$ can be chosen as the rectangular contour 
$(n-i L, n+iL, iL, -i L)$ (because $\log F_{0}(x)=\log F_{1}(x)=0$ and 
$\cot(z)$ has no poles for Im$z\neq0$). 
We can now change variable $z/n \to z$ to obtain
\be
D_n(x)=
n \int_{\cal C'}\frac{dz}{2 i} \cot(\pi z n) \log F_{z}(x)\,,
\label{Dn}
\ee
and ${\cal C'}$ is the re-scaled rectangle $(1-i L, 1+iL, iL, -i L)$.
This simple formula provides the desired analytic continuation.
Notice that all the poles of the integrand in the strip $0\leq{\rm Re}z\leq1$
are on the real axis. 
Then, if the argument of the integral would decay quickly enough
for ${\rm Im} z\to\pm\infty$,
we could send $L\to\infty$, ignoring the contribution of the horizontal pieces 
and remaining only with the vertical ones.
This is unfortunately not the case, because the integrand is increasing 
when $L\to\infty$.

\begin{figure}
\includegraphics[width=0.36\textwidth]{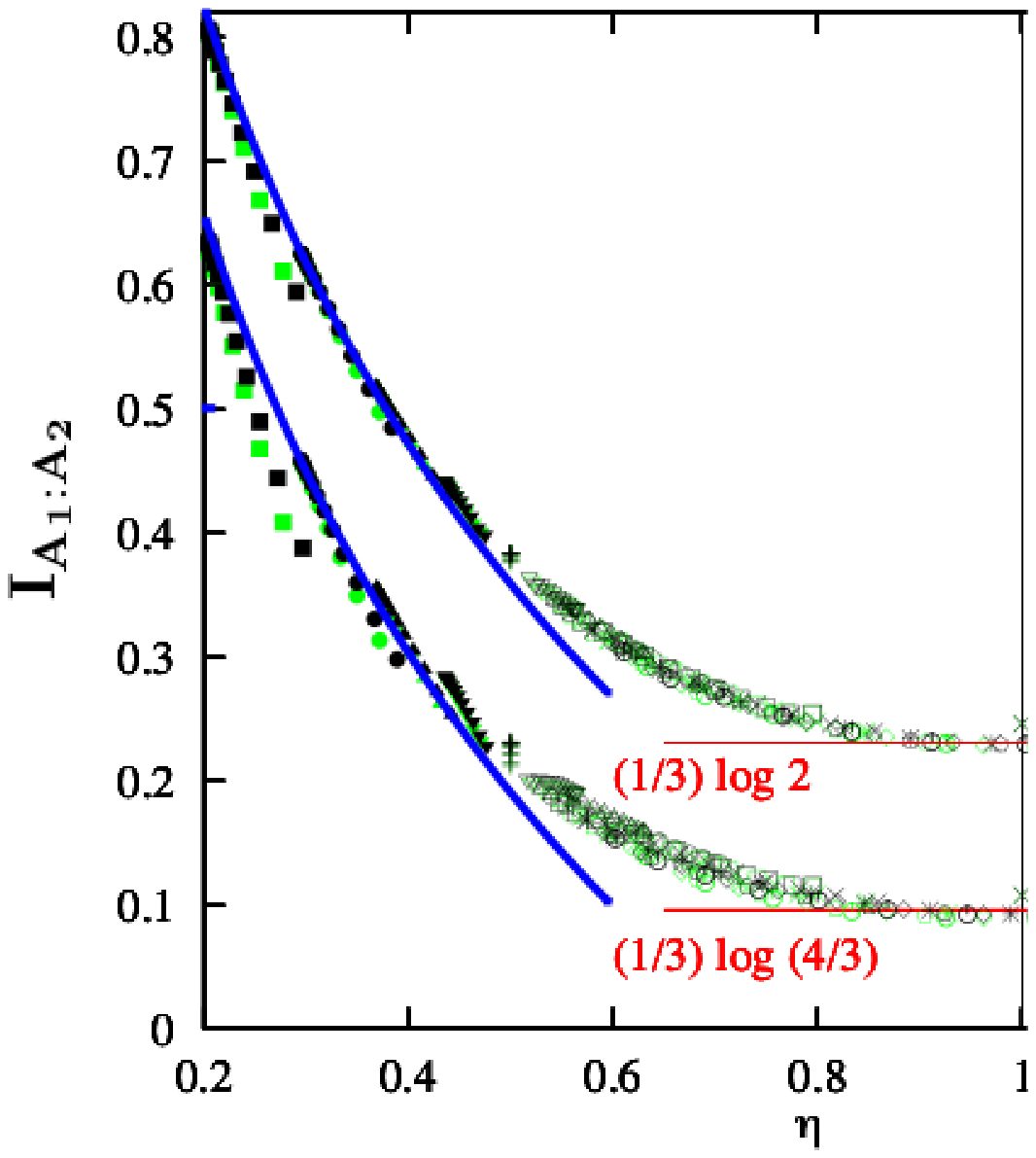}
\hspace{0.03\textwidth}
\includegraphics[width=0.58\textwidth]{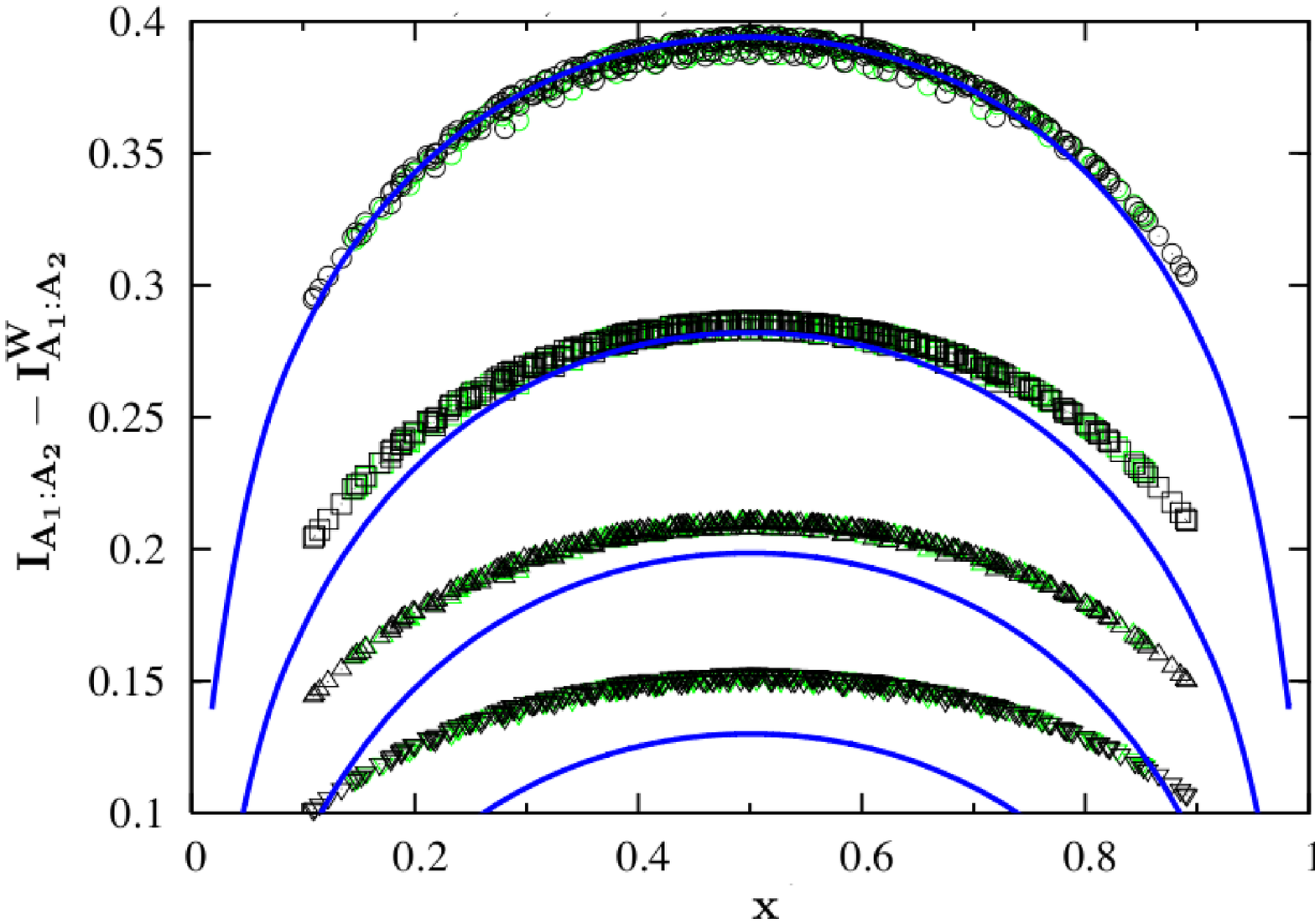}
\caption{Mutual information $I_{A_1:A_2}^{(1)}$ for the XXZ model. 
All numerical data are extracted from Ref. \cite{fps-09}.
Left: $I_{A_1:A_2}^{(1)}$ for $x=1/2$ (top curve) and $x=1/4$ (bottom curve)
as function of $\eta$. The continuous curve is the decompactification 
result Eq. (\ref{Idec}).  
Right: $I_{A_1:A_2}^{(1)}$ at fixed $\eta$ as function of $x$.
The top curve corresponds to $\eta=0.295$ small enough to agree for all 
considered $x$ with Eq. (\ref{Idec}). 
The other curves correspond to higher values of $\eta$, when 
the small $\eta$ approximation looses validity.
}
\label{Figfps}
\end{figure}

We can however take the derivative wrt $n$ in the contour integral. 
This leads to 
\be
D_1'(x)\equiv -\left.\frac{\partial D_n(x)}{\partial n}\right|_{n=1}=
\int_{\cal C'} \frac{dz}{2i} \frac{\pi z}{\sin^2 \pi z} \log F_{z}(x)\,.
\ee
In this integral, the horizontal contribution is vanishing for 
$L\to\infty$ and so only the vertical ones are left. 
Because of the periodicity of the integrand 
these two contributions are equal and so 
\be
D_1'(x)=-
\int_{-i\infty}^{i\infty} \frac{dz}{i} \frac{\pi z}{\sin^2 \pi z} \log F_{z}(x)\,.
\ee
Such integral is easily evaluated numerically for any $x$. For $x=1/2$
it is possible to get an analytic result with a different method (see 
\ref{appris}) that agrees with the value calculated numerically. 
To cross check our results in the appendix we also provide the 
analytic continuation as perturbation series in $x$. 
Despite the asymptotic character of the perturbative expansion, it provides 
a very good approximation for all $x\leq 1/2$.

\subsection{The entanglement entropy in the decompactification regime}

From Eq. (\ref{decomp}) and using the result above for the analytic 
continuation, we have that the entanglement entropy for a {\it real} 
boson in the decompactification regime is
\be
S_A(\eta\ll1)- S_A^W\simeq
\frac12 \ln \eta -\frac{D_1'(x)+D_1'(1-x)}2\,,
\ee
where $S_A^W$ is the result in Ref. \cite{cc-04}.
The same result obviously holds for $\eta\gg1$ with the replacement 
$\eta\to\eta^{-1}$.

This should be compared with the numerical results for the XXZ model by 
Furukawa et al. \cite{fps-09}, where the entanglement of the XXZ chain for 
generic values 
of the anisotropy $\Delta$ and magnetic field (always in the gapless phase) 
has been calculated by direct diagonalization for systems up to 30 spins.
In the absence of the magnetic field, $\eta$ is related to the anisotropy by
$\eta=1-(\arccos\Delta)/\pi$, while for non-zero $h_z$ a closed formula for 
$\eta$ does not exist and must be calculated numerically as 
explained in \cite{fps-09}.
The main results have been written in terms of the R\'enyi mutual informations
\be
\fl
I_{A_1:A_2}^{(n)}= S_{A_1}^{(n)}+S_{A_2}^{(n)}-S_{A_1\cup A_2}^{(n)}
= -\frac{n+1}{6n} c \log(1-x)+\frac{1}{n-1}\log \mathcal{F}_n(x)\,,
\ee
where $A_1$ and $A_2$ are the two intervals composing $A=A_1\cup A_2$.
For $n=1$ the data (reported also in Fig. \ref{Figfps}),
when plotted in terms of $x$ collapse on a single curve, confirming the validity of the scaling form. 
In Ref. \cite{fps-09}, only $I_{A_1:A_2}^{(2)}$ has been compared with the available CFT prediction Eq. (\ref{F2}). However, for $n=2$, the collapse of the data is worse than the one for $I_{A_1:A_2}^{(1)}$ because of the strong oscillating corrections to the scaling of the R\'enyi entropies (analogous to 
the ones observed for a single interval \cite{ncc-08,ccn-08}).
However, the agreement was rather satisfactory, considering the small 
system sizes and the oscillations.

\begin{figure}
\includegraphics[width=0.8\textwidth]{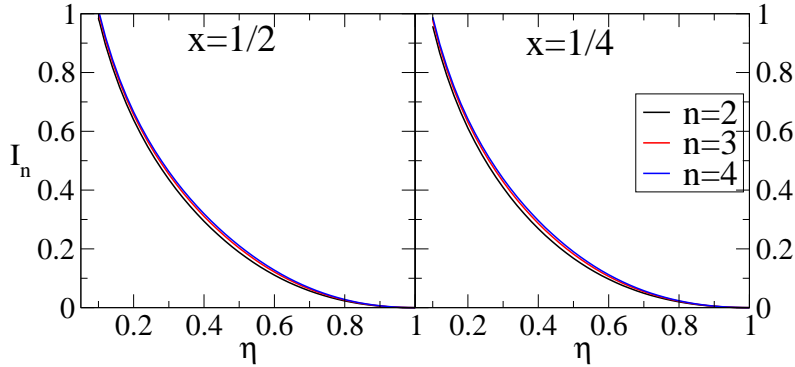}
\caption{$I_n$ in Eq. (\ref{In}) as function of $\eta$ for $x$ 
constant $x=1/2$ (left panel) and $x=1/4$ (right panel).
In each plot the three curves corresponds to different values of $n=2,3,4$ 
(from bottom to top).}
\label{Fig1}
\end{figure}

Here we are in position to offer a first prediction for the von-Neumann 
mutual information $I_{A_1:A_2}^{(1)}$. The numerical data from 
Ref. \cite{fps-09} are reported in Fig. \ref{Figfps}. 
The prediction in the decompactification regime is 
\be
I_{A_1:A_2}^{(1)}(\eta\ll1)- I_{A_1:A_2}^{(1),W}\simeq
-\frac12 \ln \eta +\frac{D_1'(x)+D_1'(1-x)}2\,,
\label{Idec}
\ee
where again $I_{A_1:A_2}^{(1),W}$ is the result of Ref. \cite{cc-04}. 
This prediction, for various values of $\eta$ and $x$ are reported in 
Fig. \ref{Figfps}. 
In the left panel it has been reported the mutual information as function
of $\eta$ at fixed $x=1/2,1/4$. On the scale of this plot, the 
decompactification prediction reproduces the numerical data up to 
$\eta\sim 0.4$ and for larger values clearly deviates.
In the right panel of Fig. \ref{Figfps} it is reported the $x$ dependence 
of the mutual information at fixed $\eta$. Again for the smallest 
value $\eta=0.295$ the decompactification approximation is valid for 
all $x$. Increasing $\eta$ this is no longer true, but it works 
better for $x\sim1/2$, while it quickly deteriorates moving from the central 
point. 

\subsection{The entanglement entropy for small $x$}

We can obtain the analytic continuation also in the small $x$ regime starting from 
the results in Sec. \ref{Secsmallx}.
We need to analytically continue ${\cal F} _n(x)$ in Eq. (\ref{Fsmallx}) valid for $\eta\neq1$. 
Calling $\alpha={\rm min} (\eta,1/\eta)$, we can write ${\cal F} _n(x)$ as
\begin{equation}
{\cal F}_n(x)= 1+ 2n\left(\frac{x}{4n^2}\right)^\alpha P_n+\dots \quad {\rm with}\quad
P_n=\sum_{l=1}^{n-1} \frac{l/n}{\left[ \sin\left(\pi{l}/{n}\right) \right]^{2\alpha}}\,.
\label{defP}
\end{equation}
Within this definition we have that, after analytically continuing, the contribution to the 
entanglement entropy is (we recall that $P_1=0$) 
\be
S_A-S_A^W=- {\cal F'}_1(x)= 2^{1-2\alpha}{x}^\alpha  P'_1+\dots\,.
\ee
The most important consequence is that for small $x$ the entanglement entropy has a power
law behavior in $x$ with an exponent that is always $\alpha$ for any $\eta\neq1$. 
The multiplicative coefficient of this power law is exactly calculated by analytically
continuing $P_n$. This derivation is however rather cumbersome and it is reported in  
\ref{appP}. The dependence of $P'_1$ from $\eta$ can be read in the plot in Fig. \ref{figP}.

\subsection{Results for integer $n$}

\begin{figure}[t]
\includegraphics[width=0.8\textwidth]{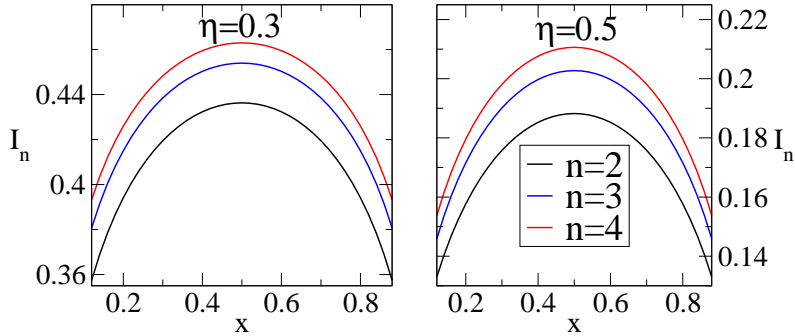}
\caption{$I_n$ in Eq. (\ref{In}) as function of $x$ at constant $\eta=0.3$ 
(left panel) and $\eta=0.5$ (right panel).
In each plot the three curves correspond to different values of $n=2,3,4$ 
(from bottom to top).
}
\label{Fig2}
\end{figure}

\begin{SCfigure}
\includegraphics[width=0.6\textwidth]{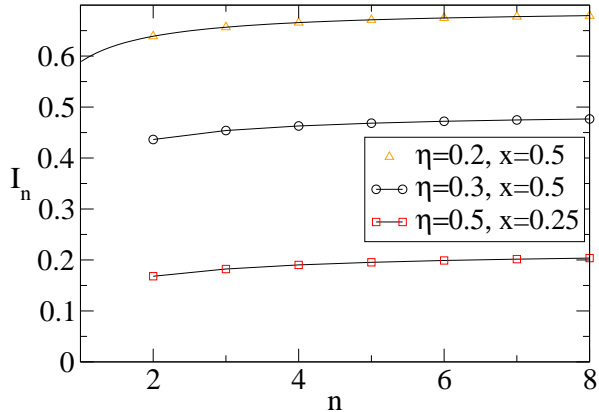}
\caption{$I_n$ in Eq. (\ref{In}) as function of $n$ at constant $\eta$ and $x$.
The points at $n$ integer are real data, the lines are only guide for the  
eyes, except the top one that is the analytic continuation for small $\eta$.   
}
\label{Fig3}
\end{SCfigure}

In this section we report some explicit results for integer $n$.
These are shown in Figs. \ref{Fig1}, \ref{Fig2}, and \ref{Fig3}, where 
we always plot the quantity
\be
I_n=\frac1{n-1}\log {\cal F}_n(x)\,,
\label{In}
\ee 
that contributes directly to the R\'enyi entropy and mutual information.
After analytic continuation, this quantity has also a smooth limit to 
$n=1$, and so it is ideal to show some general properties.

In Fig. \ref{Fig1}, following Ref. \cite{fps-09} (also left panel in 
Fig. \ref{Figfps}),  we plotted $I_n$ as function of $\eta$ with $x$ 
kept constant to $x=1/2$ (left panel) and $x=1/4$ (right panel).
In each plot the three curves correspond to different values of $n=2,3,4$.
It is evident that on the scale of the plot, the differences between 
various $n$ are tiny. This means (if nothing really strange happens 
in the analytical continuation) that also the equivalent plot 
for $I_1$ (contributing to the entanglement entropy) will be qualitatively 
and quantitatively similar. 
In particular, to appreciate the differences between various values 
of $n$ on these kinds of plots, the numerical results must be extremely precise.
In fact, the results in Ref. \cite{fps-09} (reported also in the 
left panel of Fig. \ref{Figfps}) are practically indistinguishable 
from those in Fig. \ref{Fig1}.

A more effective way to show the differences between various $n$ is 
to plot $I_n$ as a function of $x$ keeping $\eta$ constant, as in  
Fig. \ref{Fig2}. Because of the reduced scale of variation of $I_n$ 
with $x$, the differences between the various $n$ are evident and guessing 
quantitative features of the analogous plot for the analytic 
continuation at $n=1$ is not recommended. 
When $n$ increases the corresponding value of $I_n$ increases, in qualitative
agreement with the numerical results of Ref. \cite{fps-09} (see Fig. 3 there). 
As already stated, a direct comparison of the numerical and analytically 
results is quantitatively impossible because of the strong oscillating 
corrections to the scaling in the XXZ model.

Finally in Fig. \ref{Fig3} we keep constant both $\eta$ and $x$ and show 
the dependence of $I_n$ on $n$. On the scale of the plot, the $n$-dependence
is small. However, before making wrong considerations about a smooth behaviour 
in taking the limit for $n\to1$ in the analytic continuation, it is 
worth to have a look at the exact results we have for small $\eta$. 
We then consider $\eta=0.2$ and $x=1/2$ (top curve in Fig. \ref{Fig3}).
The points are the exact results for $n$ integer from Eq. (\ref{Fnv}), while 
the continuous line is the analytic continuation given 
by Eqs. (\ref{Dn12}) and (\ref{decomp}).
For such small value of $\eta$, the decompactification 
formula is indistinguishable from the exact data.
The analytic continuation displays a pronounced binding of the curve for
$1<n<2$, that makes difficult any naive extrapolation from the data with 
$n\geq2$.

\section{Conclusions}
 
We considered the entanglement of two disjoint intervals 
$A=[u_1,v_1]\cup[u_2,v_2]$ in the ground-state of the CFT of the 
Luttinger liquid (free compactified boson).
We presented a complete analysis for $\Tr \rho_A^n$ for any 
integer $n$, leading to the result we anticipated in the introduction 
Eq. (\ref{Fn}).
We have not yet been able to analytically continue the numerator of 
this expression to obtain the entanglement entropy. 
However, in the decompactification regime when the exponent $\eta$ is very small
or very large, we calculated explicitly the entanglement entropy and 
the associated mutual information leading to Eq. (\ref{Idec}). 
Our predictions agree well with the numerical computations of 
Furukawa et al. \cite{fps-09} for the XXZ model that is described by this CFT. 

The error in Ref. \cite{cc-04} also carries over to the case of 
semi-infinite systems, where now half of the points are the images ones. 
Also in this case a still unknown function $\tilde{\cal F}_n(x)$ corrects 
the result.
However, this error does not impact on the later application by two of us
of these methods to studying the time-dependence of the entanglement
entropy following a quantum quench \cite{cc-05}. This is because in
this case only the limiting behaviour, when the various
cross-ratios are either small or large, is needed, and this is
insensitive to the precise form of ${\cal F}_n(x)$. Likewise, as far as we
are aware, all the other conclusions of Ref. \cite{cc-05} remain valid.
Instead the function $\tilde{\cal F}_n(x)$ affects some results in 
Ref. \cite{cc-07l} for local quantum quenches, that however are 
easily corrected using the formulas for general four-point correlation 
functions presented in the same paper. See for a more detailed discussion 
the review \cite{cc-rev}.

The analytic continuation of ${\cal F}_n(x)$ to general real $n$ remains the 
most interesting open problem. 
However, there are several other issues that deserve to be discussed. 
Firstly it is difficult to check directly our predictions for $\Tr \rho_A^n$ 
for integer $n$ with the numerical data in Ref. \cite{fps-09}. 
This because, as observed already in \cite{fps-09} for $n=2$, there are strong 
oscillating corrections to the scaling making the comparison 
hard, if not impossible, for the system sizes accessible by exact 
diagonalization. 
There are several possible way-outs to this problem. 
One could try to describe these corrections analytically, as done for the 
single interval \cite{ccn-08} and adding them to the leading contribution,
but this seems to be very difficult.  
Alternatively one could use different numerical methods 
to access largest system sizes. 
Density matrix renormalisation group (eventually in the recent version 
proposed in Ref. \cite{Neg} to deal with a similar issue) could be effective.
Furthermore, the Monte-Carlo based approach by Caraglio and 
Gliozzi \cite{cg-08} applies to the case of integer $n$. 
This could then be used to test our predictions for models 
showing smaller oscillations (even in the same universality class).
One can also wonder whether Eq. (\ref{Fn}) is enough to calculate analytically
the full spectrum of eigenvalues of the reduced density matrix. 
In fact, in Ref. \cite{cl-08} it has been shown that this can be calculated 
by knowing $\Tr \rho_A^n$ for integer $n$ only.

A few comments are also in order. 
In Ref. \cite{ch-04,cfh-05,ch-08,ch-09,kl-08}, the 
entanglement entropy for two disjoint intervals has been calculated 
for free fermionic theories, that after bosonization always correspond 
to a compactified boson with $\eta=1/2$ \cite{cftbook}. 
However, it has been found that the entanglement entropy is given 
by $Z_{\mathcal{R}_{n,2}}^W$, apparently in contrast with the numerical 
calculation in Ref. \cite{fps-09} and what found here. 
The details of this apparent disagreement are still
not completely understood, but they should be traced back to the different boundary conditions
that result from constructing the reduced density matrix for spin or fermion variables. 
For the Ising model numerical computations \cite{ffip-08} 
also show a good agreement with $Z_{\mathcal{R}_{n,2}}^W$. Also in this case, it is
likely that the deviations from $Z_{\mathcal{R}_{n,2}}^W$ should be attributed to the 
choice of the variables used in constructing the reduced density matrix. (In fact, the 
calculations in the spin variables \cite{atc-p} show numerically and analytically 
that $Z_{\mathcal{R}_{n,2}}^W$  is not correct.) 
Finally holographic calculations in AdS/CFT correspondence \cite{rt-06,vr-08}
considering the classical limit in the gravity sector, also found 
$Z^W_{R_{n,2}}$. It would be interesting to understand how the correct 
result might arise from taking into account the quantum effects on the 
gravity side.

We close this paper by discussing the case with $N>2$ disjoint intervals. 
As far as we are aware there are no firm results in the literature. 
By global conformal invariance we have
\be
\label{generaln}
{\rm Tr}\,\rho_{A}^n=
c_n^N\left({\prod_{j<k}(u_k-u_j)(v_k-v_j)\over\prod_{j, k}(v_k-u_j)}
\right)^{(c/6)(n-1/n)}{\cal F}_{n,N}(\{ x\}) \,.
\ee
For ${\cal F}_{n,N}(\{ x\})=1$ this is the incorrect result of Ref. \cite{cc-04}
(note a typo in the denominator). $\{ x\}$ stands for the collection
of $2N-3$ independent ratios that can be built with $2N$ points.
Some old results from CFT on orbifolds in Refs. \cite{z-87,br-87}
could be useful to calculate ${\cal F}_{n,N}(\{ x\})$ for a compactified boson.

\section*{Acknowledgments}
We thank S. Furukawa, F. Gliozzi and V. Pasquier for useful discussions and correspondence.
We thank all authors of Ref. \cite{fps-09} for allowing us to use their 
numerical results.
JC thanks Benjamin Hsu for early discussions on this topic.
ET is also grateful to Hong Liu, J. McGreevy, and A. Scardicchio. 
This work was supported in part by EPSRC grant EP/D050952/1.
PC benefited of a travel grant from ESF (INSTANS activity).

\appendix

\section{Correlation functions of twist fields and ${\bf Z}_n$ orbifolds}
\label{apporb}

In a order to make this paper self-contained, in this appendix we review 
the results of Ref. \cite{Dixon} for CFT on orbifold spaces that we used.
In particular we will show how to obtain Eqs. (\ref{Zqu}) and (\ref{Scl}).
As we have stressed in the main text twist fields exist in a QFT whenever
there is a global internal symmetry. 
An orbifold is obtained by identifying points of the target space
through a given equivalence relation, leading automatically to a global 
symmetry and so to the presence of twist fields (that is why  
Ref. \cite{Dixon} contains the main ingredients for our calculations). 
To be explicit a $D$ dimensional orbifold $\mathbf{R}^D/S$ is obtained by 
identifying points of the target space ${\bf R}^D$ through an equivalence 
relation
\begin{equation}
X' \sim X\,,\qquad
{\rm if} \hspace{.8cm} X'=\theta X+v\equiv gX\,,
\hspace{.8cm} g\,=\,(\theta,v)\,,
\end{equation}
where $\theta$ is a rotation and $v$ is a vector of $\mathbf{R}^D$. 
The set of the pairs $S=\{(\theta,v)\}$ defining the orbifold is called  
{\it space group}, while the subgroup $\Lambda =\{g= (1,v)\} \subset S$, 
made by the translations only, is called {\it lattice} $\Lambda$ of $S$. 
A simple example of orbifold is obtained by identifying points in the target 
space with opposite signs $X\sim -X$. This gives rise to so-called ${\bf Z}_2$
orbifolds. 
A $\mathbf{Z}_n$ orbifold is generated by a rotation $\theta$ of order $n$ 
(i.e. $\theta^n=1$) and its space group $S$ is given by the pairs 
$(\theta^j,v)$ with $j=0,1,2, \dots, n-1$ and $v$ runs over an even 
dimensional lattice $\Lambda$.
Among the conjugacy classes making up the partition of $S$, we distinguish the
 ones of the form $\{ (1,\theta^j v_0) \}$ ($v_0$ is a fixed vector of 
$\mathbf{R}^D$), which contain only the translation elements of $S$ and 
therefore describe the winding sectors. Instead, elements like 
$(\theta^j, v)$ with $j=1,2, \dots , n-1$ and $v \in \Lambda$ belong to 
classes like
\begin{equation}
\label{conj classes}
\big\{\big(\theta^j,\theta^r v_0+(1-\theta^j) u\big); r\in \mathbf{Z} , u \in \Lambda\big\}\,,
\end{equation}
which describe the twisted sectors. Notice that, for each 
$j=1,2, \dots, n-1$, there are many conjugacy classes of the form 
$\{(\theta^j, v)\}$ with $v$ belonging to different subsets (cosets) of the 
lattice $\Lambda$.
Thus, as for the twist fields associated to the $\mathbf{Z}_n$ orbifold, 
each of them is characterized by two indices: the index $j=1, 2,\dots, ,n-1$ 
of the twisted sector of the Hilbert space and another index $\varepsilon$ 
labelling the conjugacy class within that sector.
The twist-fields in our problems are exactly the same that appears in 
${\bf Z}_n$ orbifolds, but since we do not have identifications of points 
the index $\varepsilon$ can be only zero and will be ignored in the following. 

Let us go back to our main goal, that is the calculation of the four-point
function of twist fields. 
Let us consider a complex field $X(z,\bar{z})$ defined on the worldsheet given 
by the Riemann sphere (this is one of the $\t\varphi_k$ in the main text). 
The occurrence of the  twist field  in the origin 
tells us how the field $X$ is rotated and translated when it is carried around 
this point in the worldsheet, i.e.
\begin{equation}
\label{X bc}
X(e^{2\pi i} z, e^{-2\pi i} \bar{z})\,=\,\theta^j X(z, \bar{z})+v\,,
\end{equation}
where $v$ is a vector of the coset. The phase rotation $\theta^j$  of $X$ is 
the monodromy of the field. 
In order to make this definition meaningful, the field $X$ must be complex.

\begin{figure}
\includegraphics[width=\textwidth]{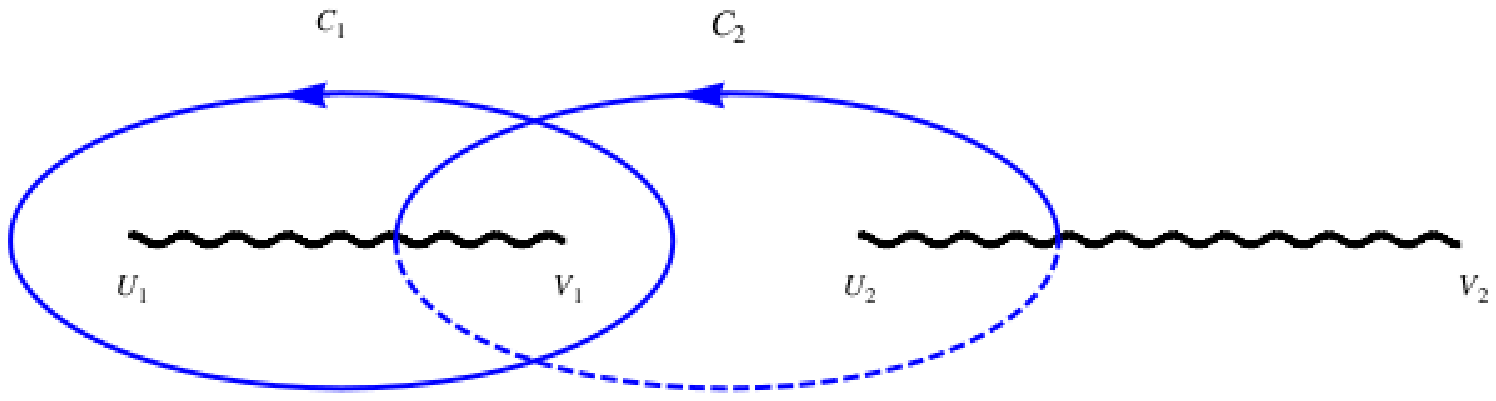}
\caption{The two closed loops ${\cal C}_1$ and ${\cal C}_2$ we considered as 
basis. The different sheets are indicated as solid vs dashed lines.}
\label{fig-C1C2}
\end{figure}

When $X$ is taken around a set of points where different twist-fields are 
placed, it is rotated and translated according to the product of the space 
group elements associated with the different twist fields. 
The relevant circuits to consider in order to completely fix the boundary 
conditions are the ones enclosing a collection of twist fields with 
{\it net twist zero}, namely the loops along which the field $X$ acquires no 
phase. Such paths $\mathcal{C}$ are called {\it closed loops}.
Two example of closed loops that we will use in the following are 
reported in Fig. \ref{fig-C1C2}.

Splitting the field $X=X_{\rm cl}+X_{\rm qu}$ into its classical part 
$X_{\rm cl}$ and its quantum counterpart $X_{\rm qu}$, we impose (\ref{X bc}) 
by requiring that
\be
\label{X bc cl}
X_{\rm cl}(e^{2\pi i} z, e^{-2\pi i} \bar{z})=\theta^j X_{\rm cl}(z,\bar{z})+v\,,
\end{equation}
and
\begin{equation}
\label{X bc qu}
X_{\rm qu}(e^{2\pi i} z, e^{-2\pi i} \bar{z})=\theta^j X_{\rm qu}(z,\bar{z})\,,
\end{equation}
namely $X_{\rm qu}$ ignores the translations in the space group. 
Thus, for any closed loop $\mathcal{C}$, we have
\begin{equation}
\label{Xqu closed loop}
\Delta_{\mathcal{C}} X_{\rm qu} =
\oint_{\mathcal{C}} dz \,\partial_z X_{\rm qu} 
+\oint_{\mathcal{C}} d\bar{z} \,\partial_{\bar{z}} X_{\rm qu} =0\,,
\end{equation}
and 
\begin{equation}
\label{Xcl closed loop}
\Delta_{\mathcal{C}} X_{\rm cl} \,=\,
\oint_{\mathcal{C}} dz \,\partial_z X_{\rm cl} 
+\oint_{\mathcal{C}} d\bar{z} \,\partial_{\bar{z}} X_{\rm cl} =v\,,
\end{equation}
where now $v$ is not the same of (\ref{X bc cl}), but it is a vector of 
$\Lambda$ depending on the twist fields enclosed by $\mathcal{C}$. 
For the twist field ${\cal T}_{k,n}$ we have 
\begin{equation}
\label{v form}
v\,\in\,(1-\theta)\Lambda\,,
\end{equation}
where, compared to Ref. \cite{Dixon}, we fixed 
$f_{\varepsilon_1}=f_{\varepsilon_2}=0$, because we have only the trivial  
fixed point $0$.
The requirements (\ref{Xqu closed loop}) and (\ref{Xcl closed loop}) 
are the {\it global monodromy conditions}.

Let us take $\theta = e^{2\pi i k/n}$ (i.e. $\theta_k$ in the main text) 
to be the rotation by an angle $2\pi k/n$ for a fixed $k\in\{1,2,\dots, n-1\}$. 
In all this appendix we will consider fixed values of $k$ and $n$, and so 
there is no ambiguity in denoting with $\tw$ and  $\t\tw$ what in the text 
we called $\tw_{k,n}$ and  $\t\tw_{k,n}$ (i.e. the twist field associated to 
$\theta$ and $\theta^{-1}$ respectively and that in \cite{Dixon} are 
called $\sigma_+$ and $\sigma_-$). We fix $j=1$ in Eqs. (\ref{X bc cl}) 
and (\ref{X bc qu}).
Also all other quantities will loose the subscripts $k$ and $n$ etc.
We are interested to the following four point function
\begin{equation}\fl
\label{4point func}
Z=
\langle \tw(z_1,\bar{z}_1)\t{\tw}(z_2,\bar{z}_2)
\tw(z_3,\bar{z}_3) \t{\tw}(z_4,\bar{z}_4)\rangle
=\int [dX] [d\bar{X}] e^{-S[ X,\bar{X} ]}\,,
\end{equation}
where the action for the complex field $X$ reads
\begin{equation}
S[X,\bar{X}]=\frac{1}{4\pi} \int
\big(\partial_z X \partial_{\bar{z}} \bar{X} 
+ \partial_{\bar{z}} X \partial_z \bar{X} \big) \,d^2z\,.
\end{equation}
Separating the classical contribution from the quantum part, 
we can write
\begin{equation}
\label{Z=Zcl Zqu}
Z
=Z_{\rm qu} \sum_{\langle X_{\rm cl} \rangle} e^{-S_{\rm cl}}\,,
\end{equation}
where $S_{\rm cl} \equiv S[X_{\rm cl},\bar{X}_{\rm cl}]$. 
The classical part in (\ref{Z=Zcl Zqu}) is given by the sum over the possible 
configurations of the classical field, which are characterized 
by (\ref{Xcl closed loop}). 

Following \cite{Dixon}, let us start by considering the Green function
in the presence of four twist-fields
\be
g(z,w;z_i)=\frac{\langle-\frac{1}{2} \partial_z X \partial_w \bar X
\tw(z_1)\t{\tw}(z_2)\tw(z_3)\t{\tw}(z_4)\rangle}
{\langle \tw(z_1)\t{\tw}(z_2)\tw(z_3)\t{\tw}(z_4)\rangle}\,.
\ee
Imposing that for $z\to w$ we have $g(z,w;z_j)\sim (z-w)^{-2}$ and that 
for $z\to z_j$ we have $g(z,w;z_j)\sim (z-z_j)^{-k/n}$
and  $g(z,w;z_j)\sim (z-z_j)^{-(1-k/n)}$ for $j$ odd and even respectively (and 
the opposite for $w\to z_j$), we can write $g(z,w;z_j)$ as 
\bea\fl
g(z,w;z_i)&=&\omega_k(z)\omega_{n-k}(w) \left[
\frac{k}n \frac{(z-z_1)(z-z_3) (w-z_2)(w-z_4)}{(z-w)^2}+
\right.\nonumber\\\fl&&\qquad\left.
\left(1-\frac{k}n\right) 
\frac{(z-z_2)(z-z_4) (w-z_1)(w-z_3)}{(z-w)^2}+A(z_j,{\bar z}_j)\right]\,,
\eea
where
\be
\omega_k(z)=[(z-z_1)(z-z_3)]^{-k/n} [(z-z_2)(z-z_4)]^{-(1-k/n)}\,,
\label{omegak}
\ee
and $A(z_j,{\bar z}_j)$ is a constant (in $z$ and $w$) that contains the 
dependence of $g(z,w;z_i)$ on the antiholomorphic coordinates and must 
be determined by global monodromy conditions.

Let us now consider the limit $w\to z$ 
\bea\fl
\lim_{w\to z}[g(z,w)- (z-w)^{-2}]&=&
\frac12\frac{k}n\left(1-\frac{k}n\right) 
\left(\frac1{z-z_1}+\frac1{z-z_3}-\frac1{z-z_2}-\frac1{z-z_3}\right)^2+
\nonumber\\\fl&&\qquad
+\frac{A(z_j,{\bar z}_j)}{(z-z_1)(z-z_2)(z-z_3)(z-z_4)}\,.
\label{gwz}
\eea
This is exactly the expectation value of the insertion of the 
stress energy tensor of the field $X$ in the four-point correlation function. 
By taking the limit $z\to z_j$ for any $j$, one reads the dimensions of the 
twist fields $\tw$ and $\t\tw$
\begin{equation}
\Delta_{\frac{k}{n}} = \bar{\Delta}_{\frac{k}{n}}  
=\frac{1}{2}\frac{k}{n}\left(1-\frac{k}{n}\,\right)\,,
\end{equation}
as anticipated in the main text.
Let us also introduce the auxiliary correlation function
$$
h(\bar z,w;z_j)=
\frac{\langle -\frac{1}{2}\partial_{\bar{z}}X \partial_w \bar{X}
\tw(z_1)\t{\tw}(z_2)\tw(z_3)\t{\tw}(z_4)\rangle}{\langle\tw(z_1)\t{\tw}(z_2)\tw(z_3)\t{\tw}(z_4)\rangle}=
B(z_j,\bar z_j) \bar \omega_{n-k}(\bar z) \omega_{n-k}(w)\,,
$$
where the rhs above comes from the same considerations as before, but the 
singular terms in $(z-w)$ do not occur because $h(\bar z,w;z_j)$ is regular 
in this variable.

In order to lighten the notation it is convenient hereafter to employ the following conformal
map
\be
z \rightarrow \frac{(z_1-z)(z_3-z_4)}{(z_1-z_3)(z-z_4)}
\ee 
which sends $z_1$, $z_2$, $z_3$ and $z_4$ into $0$, $x$, $1$ and $\infty$
respectively, where $x$ is the four-point ratio ($z_{ij}=z_i-z_j$)
\begin{equation}
x\equiv\frac{z_{12} \,z_{34}}{z_{13} \, z_{24}}, \qquad 
\label{xxx relation}
x(1-x)=\frac{z_{21}\, z_{43}\, z_{41}\, z_{32}}{z_{31}^2\, z_{42}^2}\,.
\end{equation}
After this mapping the dependence on $z_i$ in the varius functions reduces to the 
dependence on the ratio $x$ and its complex conjugate $\bar{x}$ only.

Taking the limit $z\to x$ in Eq. (\ref{gwz}), one finds
the following differential equation for the quantum part of the 
correlation function
\be
\partial_x \ln Z_{\rm qu}(x,\bar x) =-2\Delta_{{k}/n} 
\left(\frac1x-\frac1{1-x}\right)-\frac{A(x,\bar x)}{x(1-x)}\,,
\ee

The global monodromy conditions allow to determine $A(x,\bar x)$.
In terms of the functions $g$ and $h$ they read
\be
0=\oint_{\cal C} dz g(z,w)+\oint_{\cal C} d\bar z h(\bar z,w)\,.
\ee
Dividing by $\omega_{n-k}(w)$ and letting $w\to\infty$ this gives
\be\fl
A(x,\bar x) \oint_{\cal C} dz \omega_k(z)+ 
B(x,\bar x) \oint_{\cal C} d\bar z \bar \omega_k(\bar z)=
-\left(1-\frac{k}n\right)\oint_{\cal C} dz (z-x) \omega_k(z)\,.
\label{mon}
\ee
These integrals are valid for all closed loops. We choose as a basis
of these loops the two ones depicted in Fig. \ref{fig-C1C2}, that  
suffices to determine $A(x,\bar x)$.

All these integrals are easily calculated giving \cite{Dixon}
\bea\fl &&
\label{A24}
\oint_{{\cal C}_1} dz \omega_k(z)= 2\pi i e^{-i\pi k/n} F(x),\qquad
\oint_{{\cal C}_2} dz \omega_k(z)= 2\pi i  F(1-x),\\ \fl &&
\oint_{{\cal C}_1} d\bar z \bar \omega_{n-k}(\bar z)= 2\pi i e^{-i\pi k/n} \bar F(\bar x),\qquad
\oint_{{\cal C}_2} d\bar z \bar \omega_{n-k}(\bar z)= -2\pi i  \bar F(1-\bar x),\\ \fl &&
-\left(1-\frac{k}n\right)\oint_{{\cal C}_1} dz (z-x) \omega_k(z)=
  2\pi i e^{-i\pi k/n} x(1-x) \frac{d F(x)}{dx}\,,\\ \fl &&
-\left(1-\frac{k}n\right)\oint_{{\cal C}_2} dz (z-x) \omega_k(z)=
  2\pi i  x(1-x) \frac{d F(x)}{dx}\,,
\eea
where we introduced
\begin{equation}
F(x)\,\equiv\, _2 F_1(k/n,1-k/n;1;x)\,.
\end{equation}
From the two equations (\ref{mon}) for ${\cal C}_1$ and  ${\cal C}_2$, 
eliminating $B(x,\bar x)$ and solving for $A(x,\bar x)$ we get 
\be
A(x,\bar x)=x(1-x) \frac{d \ln I(x,\bar x)}{dx}\,,
\ee
with 
\begin{equation}
I(x, \bar{x})\equiv
F(x) \bar{F}(1-\bar{x})+\bar{F}(\bar{x}) F(1-x)
=2 \beta(x) |F(x)|^2\,,
\end{equation}
and we also introduced 
\begin{equation}
\label{tau}
\tau(x)\,\equiv\,\alpha(x)+ i\,\beta(x)\equiv\,i\,\frac{F(1-x)}{F(x)}.
\end{equation}
For $n=2$, $\tau(x)$ gives the modulus of the torus, which is the covering 
surface of our Riemann sphere with four branch points, but for higher values 
of $n$ this is not true anymore and (\ref{tau}) has no geometric meaning.
$A(x,\bar x)$ gives the desired quantum partition function
\begin{equation}
\label{Zqu 2int}
Z_{\rm qu}(x, \bar{x})\,=\,
\frac{\rm const}{|x(1-x)|^{4\Delta_{k/n}}}\frac{1}{I(x, \bar{x})}\,.
\end{equation}
When this equation is specialized to $x$ real, it gives exactly the result 
anticipated in the main text in Eq. (\ref{Zqu}).

For the classical part, we only need to construct the properly normalised 
classical solutions. From the equations of motion, one easily sees that 
$\partial_z X_{\rm cl}$ and $\partial_z \bar{X}_{\rm cl}$ are holomorphic 
while $\partial_{\bar{z}} X_{\rm cl}$ and $\partial_{\bar{z}} \bar{X}_{\rm cl}$ 
are antiholomorphic and they can be written as
\begin{equation}\fl
\label{ab defs}
\begin{array}{lll}
\partial_z X_{\rm cl} (z) \,=\,a\,\omega_k(z)\,, 
& \hspace{1.5cm}& \partial_{\bar{z}} X_{\rm cl} (\bar{z}) \,=\,b\,\bar{\omega}_{n-k}(\bar{z})\,,  \\
\rule{0pt}{.5cm}
\partial_z \bar{X}_{\rm cl} (z) \,=\,\tilde{a}\,\omega_{n-k}(z)\,, 
& \hspace{1.5cm}& \partial_{\bar{z}} \bar{X}_{\rm cl} (\bar{z}) \,=\,\tilde{b}\,\bar{\omega}_{k}(\bar{z})\,,
\end{array}
\end{equation}
where $\omega_k(z)$ is given in Eq. (\ref{omegak}).

The complex constants $a$, $\tilde{a}$, $b$ and $\tilde{b}$ are fixed 
through the global monodromy conditions (\ref{Xcl closed loop}) for the 
closed loops $\mathcal{C}_1$ and $\mathcal{C}_2$. 
(We do not adopt here the notation of \cite{Dixon}, where 
$\bar{a} \equiv \tilde{a}$ and $\bar{b} \equiv \tilde{b}$, because we find it 
misleading.)
In order to write them, one constructs two classical solutions $X_{\rm cl,1}$ 
and $X_{\rm cl,2}$ having the following simple global monodromy conditions
\begin{equation}
\Delta_{\mathcal{C}_i} X_{{\rm cl},j}
\,=\, \Delta_{\mathcal{C}_i} \bar{X}_{{\rm cl},j}
\,=\,2\pi\,\delta_{ij}\,,
\hspace{1.6cm} i,j\,=\,1,2\,,
\end{equation}
and finds out the corresponding complex constants $a_i$, $\tilde{a}_i$, $b_i$ 
and $\tilde{b}_i$, which read
\begin{eqnarray} \fl
a_1=-e^{2\pi i \frac{k}{n}}\tilde{a}_1
=-i e^{\pi i \frac{k}{n}}\frac{\bar{F}(1-\bar{x})}{I(x,\bar{x})}\,,
& \hspace{.8cm}&
a_2=\tilde{a}_2
=-i \frac{\bar{F}(\bar{x})}{I(x,\bar{x})}\,,
\hspace{1.5cm}\\ \fl
\rule{0pt}{.8cm}
b_1=-e^{2\pi i \frac{k}{n}}\tilde{b}_1
=-i \,e^{\pi i \frac{k}{n}}\frac{F(1-x)}{I(x,\bar{x})}\,,
& &
b_2=\tilde{b}_2
=+ i \frac{F(x)}{I(x,\bar{x})}\,.
\end{eqnarray}
Then, from (\ref{Xcl closed loop}) for $\mathcal{C}_1$ and $\mathcal{C}_2$ 
and (\ref{v form}), one gets the complex coefficients for $X_{\rm cl}$ to 
use in (\ref{ab defs}), which read
\begin{eqnarray} 
a\,=\, a_1 v_1+a_2  v_2\,, 
& \hspace{1cm}&  b\,=\, b_1 v_1+b_2 v_2\,,\\
\tilde{a}\,=\, \tilde{a}_1 \bar{v}_1+ \tilde{a}_2 \bar{v}_2\,,
& & \tilde{b}\,=\, \tilde{b}_1 \bar{v}_1+ \tilde{b}_2 \bar{v}_2\,,
\end{eqnarray}
with the vectors characterizing the global monodromy condition of 
$X_{{\rm cl},1}$ and $X_{{\rm cl},2}$ given by
\begin{equation}
v_{1,2}\in(1-\theta)\Lambda\,.
\end{equation}
The last step we need for our purposes is the expression for $S_{\rm cl}$. 
By employing the following integral
\begin{equation}\fl
\label{Dixon CP1 integral}
\int |\omega_k|^2 d^2z = \int
\frac{d^2z}{|z|^{2\frac{k}{n}} |z-x|^{2\left(1-\frac{k}{n}\right)} |z-1|^{2\frac{k}{n}} }
=\frac{\pi^2}{\sin(\pi {k}/{n})} I(x,\bar{x})\,,
\end{equation}
one finds that
\begin{equation}\fl
\label{dixon Scl}
S_{\rm cl}(v_1,v_2)=
\frac{\pi \sin(\pi k/n)}{\beta}\,
  \Big[|\xi_1|^2|\tau|^2+\alpha \big(\xi_1\bar{\xi}_2 \bar{\gamma}+\bar{\xi}_1\xi_2 \gamma\big)
  +|\xi_2|^2\Big]\,,
\end{equation}
where $\gamma \equiv -i e^{-i\pi{k}/{n}}$ and we introduced the vectors 
$\xi_j$ ($j=1,2$) independent of $k/n$ that are
generic vectors of the target space lattice $\Lambda$, 
having therefore $v_j = (1-\theta)\xi_j$.

When we specialize to the problem with real $x  \in (0,1)$, 
some simplifications occur in the formulas. 
$\tau$ is purely imaginary (i.e. $\alpha=0$) and $I(x,x)=2F(x) F(1-x)$. 
Then Eq. (\ref{dixon Scl}) reduces to Eq. (\ref{Scl}) in the main text, 
where we restored all the $k$ and $n$ dependence in each quantity.

\section{A transformation formula for the Riemann-Siegel theta function}
\label{apppoi}

The Riemann-Siegel theta function $\Theta(0|T)$ for any symmetric complex 
matrix $T$ with positive imaginary part can be re-written as
\begin{equation}
\label{theta def delta}
\Theta(0|T)\,=\,
\sum_{m\,\in\,\mathbf{Z}^g} e^{i\pi\,m^{\rm t}\cdot  T\cdot m}
=\int d^gs \sum_{m\,\in\,\mathbf{Z}^g}\delta_g(s-m) \; e^{i\pi\,s^{\rm t} \cdot T \cdot s}\,.
\end{equation}
Now we employ the following identity (which is a special case of 
the Poisson resummation formula)
\begin{equation}
\label{sum delta}
 \sum_{m\,\in\,\mathbf{Z}^g}\delta_g(s-m) 
 \,=\,
  \sum_{j\,\in\,\mathbf{Z}^g} e^{2\pi i \,j^{\rm t} \cdot s}\,.
\end{equation}
Plugging (\ref{sum delta}) into (\ref{theta def delta}) and inverting $\int d^g s$ and $\sum_{j \in\mathbf{Z}^g}$ in (\ref{theta def delta}), we get a $g$ dimensional gaussian integral, which gives 
\begin{equation}
\Theta(0|T)\,=\,
\frac{\Theta(0|-T^{-1})}{\sqrt{{\rm det}(-\,i\,T)}}\,.
\end{equation}
Now, by applying this formula for $T=\lambda \,\t{\Gamma} = -A^{\rm t} (\Gamma/\lambda)^{-1} A$, where $A \equiv 2\,\widehat{U}^\dagger H \,\widehat{U}$ with $H = \textrm{diag} (\dots , \sin(\pi k/n) , \dots)$, we get
\begin{equation}
\Theta(0|\eta \,\t{\Gamma})\,=\,
\sqrt{\textrm{det}\big(\Gamma/(i\eta)\big)}\,
\Theta(0| A^{-1}(\Gamma/\eta)A^{-1} )\,,
\end{equation}
where we have used that $A^{\rm t} =A$ and $\textrm{det}\,A=1$ (here $A$ and 
$\Gamma$ are the matrix defined in the main text). 
We employ the following identity
\begin{equation}
\Theta(0| A^{-1}(\Gamma/\eta)A^{-1} ) = \Theta(0| \Gamma/\eta)\,, 
\label{boh}
\end{equation}
which is numerically true, but we give here without a proof.
Putting everything together we have 
\begin{equation}
\Theta(0|\eta \,\t{\Gamma} )\,=\,
\frac{1}{\eta^{\frac{n-1}{2}}}\left(\prod_{k=1}^{n-1} \beta_{k/n}\right)^{1/2}
 \Theta(0| \Gamma/\eta )\,, 
\end{equation}
which is exactly Eq. (\ref{theta gamma tilde}) we wanted to prove.

\section{A Thomae-type formula for singular $\mathbf{Z}_n$ curves}
\label{apptom}

In this appendix we employ a Thomae-type formula obtained in 
\cite{EnolskiGrava} for singular $\mathbf{Z}_n$ curves to prove 
equation (\ref{th1}).
In \cite{EnolskiGrava}, the following singular $\mathbf{Z}_n$ curves 
$\mathcal{C}_{n,m}$ 
are considered
\begin{equation}\fl
w^n \,=\,p(z)\,q(z)^{n-1}\,,
\hspace{1.2cm}
p(z)\,\equiv\,\prod_{j\,=\,0}^{m} (z-z_{2j+1})\,,
\hspace{.7cm}
q(z)\,\equiv\,
\prod_{j\,=\,1}^{m} (z-z_{2j})\,.
\end{equation}
They have singularities at the points 
$P_2=(z_{2},0)$, $\dots$, $P_{2m}=(z_{2m},0)$, while 
$P_1=(z_{1},0)$, $\dots$, $P_{2m+1}=(z_{2m+1},0)$, 
$P_{2m+2}=P_{\infty}=(\infty,\infty)$ are the branch points. 
These curves are $n$ sheeted coverings of the complex plane 
and they define Riemann surfaces of genus $(n-1)m$.
The cycles $\alpha_{k}$ and $\beta_k$ ($k=1, \dots , (n-1)m$) (generalizations
of ${\cal C}_1$ and ${\cal C}_2$ in Fig. \ref{fig-C1C2} for the many 
intervals situation, see Ref. \cite{EnolskiGrava} for a pictorial 
representation) provide the basis of the closed loops. 
We also introduce 
\begin{equation}\fl
\label{u diff}
du_{j+(k-1)m}\,\equiv\,\frac{z^{j-1} \,q(z)^{k-1}}{\mu^k}\,dz\,,
\hspace{1.5cm}
j\,=\,1, \dots , m\,, \hspace{.6cm} k\,=\,1, \dots , n-1\,,
\end{equation}
which are a basis of the canonical holomorphic differentials. 
The $(n-1)m \times (n-1)m$ matrices ${\cal A}$ of the $\alpha$-periods 
and ${\cal B}$ of the $\beta$-periods, whose elements read
\begin{equation}\fl
{\cal A}_{st} \,\equiv\, \oint_{\alpha_s} du_{t}\,,
\hspace{1.2cm}
{\cal B}_{st} \,\equiv\, \oint_{\beta_s} du_{t}\,,
\hspace{1.2cm}
s,t\,=\,1, \dots, (n-1)m\,,
\end{equation}
respectively, can be written in terms of the following $m \times m$ 
matrices ($k = 1, \dots , n-1$)
\begin{equation}\fl
({\cal A}_k)_{ij} \,\equiv\, \oint_{\alpha_i} du_{j+m(k-1)}\,,
\hspace{.8cm}
({\cal B}_k)_{ij} \,\equiv\, \oint_{\beta_i} du_{j+m(k-1)}\,,
\hspace{0.8cm}
i,j \,=\,1, \dots, m\,,
\end{equation}
as follows
$$
{\cal A}=\textrm{diag} ({\cal A}_1, \dots , {\cal A}_{n-1}){\cal R}_{\cal A}\,,
\hspace{1.2cm}
{\cal B}=\textrm{diag} ({\cal B}_1, \dots , {\cal B}_{n-1}){\cal R}_{\cal B}\,,
$$
where ${\rm diag} ({\cal A}_1, \dots , {\cal A}_{n-1})$ and 
${\rm diag} ({\cal B}_1, \dots , {\cal B}_{n-1})$ are block diagonal. 
The matrices ${\cal R}_{\cal A}$ and ${\cal R}_{\cal B}$ are defined as 
${\cal R}_{\cal A} = \t{\cal R}_{\cal A} \otimes \textrm{id}_m$ and 
${\cal R}_{\cal B} = \t{\cal R}_{\cal B} \otimes \textrm{id}_m$ 
($\textrm{id}_m$ is the $m \times m$ identity matrix), where the 
$(n-1)\times (n-1)$ matrices $\t{\cal R}_{\cal A}$ and 
$\t{\cal R}_{\cal B}$ have the following elements
\begin{equation}\fl
(\t{\cal R}_{\cal A})_{kr} \equiv \rho^{k(1-r)}\,,
\hspace{1.2cm}
(\t{\cal R}_{\cal B})_{kr} \equiv  \frac{\rho^k}{1-\rho^k}(\rho^{-kr}-1)\,,
\hspace{1.5cm}
\rho\,\equiv\,e^{\frac{2\pi i}{n}}\,.
\end{equation}
Introducing the normalised holomorphic differentials 
$d\vec{v} =(dv_1, \dots , dv_{(n-1)m}) = d\vec{u} \,{\cal A}^{-1}$, 
allows to define the Riemann period matrix as
\begin{equation}
\Pi_{st}\,\equiv\,\oint_{\beta_s} dv_t\,,
\hspace{2cm}
s,t\,=\,1, \dots , (n-1)m\,,
\end{equation}
which turns out to be
\begin{equation}
\label{pi matrix}
\Pi={\cal R}_{\cal A}^{-1}\, 
\textrm{diag}({\cal A}_1^{-1} {\cal B}_1,\dots,{\cal A}_{n-1}^{-1}{\cal B}_{n-1})
{\cal R}_{\cal B}\,. 
\end{equation}
Given these definitions, in \cite{EnolskiGrava} the following 
{\it Thomae-type formula} (see (5.28) of \cite{EnolskiGrava}) has been proven
\begin{equation}\fl
\label{thomae}
\Theta^8 (0|\Pi)=
\prod_{k=1}^{n-1}
\left[\frac{\textrm{det} {\cal A}_k}{(2\pi i)^m }\right]^4
\prod_{1\leq i<j\leq m}
\big(z_{2i}-z_{2j}\big)^{2(n-1)}\hspace{-2mm}
\prod_{0\leq i<j\leq m}\hspace{-2mm}
\big(z_{2i+1}-z_{2j+1}\big)^{2(n-1)}\,.
\end{equation}

In our case, we have $m=1$ i.e. four points  $z_1,z_2,z_3$, and $z_4$;
thus $p(z)=(z-z_1)(z-z_3)$ and $q(z)=(z-z_2)(z-z_4)$. 
In the rhs of (\ref{thomae}), the product of $(z_{2i}-z_{2j})$ is $1$ 
because $1 \leq i<j \leq 1$ cannot be fulfilled, while the product of 
$(z_{2i+1}-z_{2j+1})$ has only one term, which is $1$.
The two independent cycles $\alpha_1$ and $\beta_1$ coincide 
respectively with ${\cal C}_1$ and ${\cal C}_2$ in Fig. \ref{fig-C1C2}. 
The matrices ${\cal A}_k$ are one-by-one, and therefore 
$$
\Pi = {\cal R}_{{\cal A}} ^{-1} \,
\textrm{diag} ( {\cal B}_1/ {\cal A}_1, \dots , {\cal B}_{n-1}/ {\cal A}_{n-1})\,
{\cal R}_{{\cal B}}\,. 
$$ 
Moreover, ${\cal R}_{\cal A} = \t{\cal R}_{\cal A}$ 
and ${\cal R}_{{\cal B}} = \t{{\cal R}}_{{\cal B}}$ 
and (\ref{u diff}) reduces to $du_{k}(z)=\omega_k(z) dz$.
Thus, we have that ${\cal A}_k$ and ${\cal B}_k$ are given by the integrals 
in Eqs. (\ref{A24}). 
We have the following expression for $\Pi$
\begin{equation}
\label{Pi matrix}
\Pi \,=\, \t{{\cal R}}_{{\cal A}} ^{-1} \,
\textrm{diag} (\dots , \rho^{k/2} \beta_{k/n}\, , \dots)\,
\t{{\cal R}}_{{\cal B}} \,.
\end{equation}
By introducing the $(n-1)\times (n-1)$ matrix $M_1$ having all the elements 
equal to 1, we can 
write $\t{{\cal R}}_{{\cal A}} ^{-1}  $ and $\t{{\cal R}}_{{\cal B}} $ 
in terms of the matrix $\widehat{U}$ as follows
\begin{eqnarray}
\t{{\cal R}}_{{\cal A}} ^{-1}  
&=&\frac{1}{\sqrt{n}} 
\big(\textrm{id}_{n-1}+M_1\big)\,\widehat{U}\;
\textrm{diag} (\dots , \rho^{-k} , \dots)\,,
\\
\rule{0pt}{.7cm}
\t{{\cal R}}_{{\cal B}} 
&=&\sqrt{n}\;
\textrm{diag} \left(\dots , \frac{\rho^{k}}{1-\rho^k} , \dots\right)
\widehat{U}^\dagger \big(\textrm{id}_{n-1}+M_1\big)\,.
\end{eqnarray}
Now we observe that the matrix (\ref{Pi matrix}) is related to $\Gamma$ 
of Eq. (\ref{Gammadef}) as
\begin{equation}
\Pi=A^{-1}\, \Gamma A^{-1}\,,
\end{equation}
where $A \equiv 2\,\widehat{U}^\dagger H \,\widehat{U}$ with $H = \textrm{diag} (\dots , \sin(\pi k/n) , \dots)$.
As for the rhs of (\ref{thomae}), it reduces to (notice that 
$\prod_{k=1}^{n-1} \rho^{k/2}=(-i)^{n-1}$)
\begin{equation}
\prod_{k\,=\,1}^{n-1}
\left(\frac{{\cal A}_k}{2\pi }\right)^4
=\left(\,\prod_{k\,=\,1}^{n-1} \,
 _2F_1\big(k/n,1-k/n;1;x\big)\right)^4\,.
\end{equation}
Using finally $\Theta(0|A^{-1}\, \Gamma A^{-1}) = \Theta(0|\Gamma)$ 
Eq. (\ref{boh}), we have therefore proved (\ref{th1}).

\section{Three analytic continuations more}
\label{appris}

\subsection{The analytic case $x=1/2$}

For $x=1/2$, the hypergeometric function can be written in terms 
of $\Gamma$ function as
\be
F_{k/n}(1/2)=
\frac{\sqrt{\pi}}{\Gamma(1-k/(2 n)) \Gamma (k/(2 n)+1/2)}\,,
\ee
so that the $D_n$ simplifies to
\be
D_n(1/2)=\log \prod_{k=1}^{n-1}  F_{k/n}(1/2)=
\log \frac{\pi^{(n-1)/2}}{\prod_{l=1}^{n-1} [\Gamma(1/2+l/(2n))]^2}\,.
\ee
We can then employ the following integral representation of the logarithm 
of the $\Gamma$ function
\be
\log \Gamma(z) =\int_0^\infty\frac{dt e^{-t}}{t}\left( 
\frac{e^{-(z-1) t} - 1}{1 - e^{-t}} + z - 1\right)\,,
\ee
and so 
\be\fl
\log \Gamma\left(\frac12+\frac{l}{2n}\right)=
\int_0^\infty\frac{dt e^{-t}}{t}\left( 
\frac{e^{-(l/(2n)-1/2) t} - 1}{1 - e^{-t}} + \frac{l}{2n}-\frac12 \right)\,,
\ee
The resulting series is easily summed
\be\fl
D_n(1/2)=\frac{n-1}2\ln\pi-2 \int_0^\infty\frac{dt e^{-t}}{t}\left(
\frac{e^t(e^{t/2}-1-n(e^{t/(2n)}-1))}{(e^t-1)(e^{t/(2n)}-1)}+\frac{1-n}4
\right)\,,
\label{Dn12}
\ee
and the derivative wrt $n$ taken
\be\fl
D_1'(1/2)=-\frac12\ln\pi+\int_0^\infty \frac{dt e^{-t}}{2t} 
\frac{e^{3 t/2} (2 t-5)+5 e^t+e^{t/2}-1}{(e^t-1)(e^{t/2}-1)}\,.
\ee
The integral is convergent, but the various pieces in which it can be divided 
are not and so a lot of care must be used to perform it. 
The easiest way we find out to make the integral is to make a sort 
of dimensional regularization of the term $t^{-1}\to t^{-1+\nu}$, making then
the various integrals that are finite for large enough $\nu$ and then 
expanding the result close to $\nu=0$ where all the divergences cancel 
leaving (after long algebra) the finite result
\be
D_1'(1/2)=\frac{1}{2} (-1-\gamma_E +\ln \pi )\simeq -0.216243\,,
\ee
where $\gamma_E$ is the Euler $\gamma$ constant.

\subsection{Perturbative expansion in $x$}

The easiest way to perform the analytic continuation is to expand in power of
$x$ the function $F_{k/n}(x)$. From the well-known series of the hypergeometric
function the coefficient of this expansion at order $p$ is a polynomial 
of order $2p$ in $k/n$. The product over $k$ from $1$ to $n-1$ can be 
done for very large orders and the derivative for $n=1$ taken.
With a little help from Mathematica we obtain the following perturbative series
for $D_1'(x)$:\footnote{A numerology digression: 
The denominators in this Taylor series are integer multiple of the 
ones of $\tan x$ up to order $10$ when $D_1'(x)$ starts oscillating. 
On the Sloane on-line encyclopedia of integer sequences, another 
``look-a-like the denominators in Taylor series for $\tan x$'' can be found, 
(sequence number A156769) that starts differentiating from $\tan x$ at the 
eleventh term (excluding $x$), exactly like this one.}  
\bea\fl
-D_1'(x)&=&
\frac{x}{3} + \frac{2}{15}x^2 + \frac{26}{315}x^3 + \frac{2}{35}x^4 
+ \frac{52}{1155}x^5 + \frac{302}{9009}x^6 + \frac{76}{2145}x^7 + 
\frac{398}{255255}x^8 
\nonumber \\ \fl&& \qquad
+ \frac{327128}{2297295}x^9 - 
\frac{18047684}{24249225}  x^{10} +\frac{31378136}{5311735}x^{11}+O(x^{12})\,.
\eea
It is easy to obtain as many terms of this series as we want, but this is 
useless if we perform a direct summation. 
In fact, it is evident that the terms up $x^8$ are all positive and 
quickly decreasing in magnitude, giving the impression of a convergent 
series. Unfortunately from the term $x^9$ they start increasing abruptly and 
oscillating signaling that $D_1'(x)$ is an asymptotic series. 
It would be possible to resum this series a'la Borel, but this is of 
no use in view of the exact result we obtained in the main text. 
This result, that is very easy to obtain, has been a fundamental cross-check 
for other relations we found. It also provides a very good 
estimation of $D_1'(x)$ for $x\leq 1/2$. 
For example, summing the first eight terms for $x=1/2$ we obtain
$D_1'(1/2)\sim -0.216102$, deviating only of the $0.06\%$ from the exact result 
above.

\subsection{Analytic continuation for small $x$}
\label{appP}

In this appendix we report the details of the analytic continuation in the small $x$ regime. 
We need to continue to general complex $n$ the sum $P_n$ in Eq. (\ref{defP}).
It is instructive to consider first two special cases that are easily worked out. 
For $\alpha=1$ we have
\begin{equation}
P_n=
\sum_{l=1}^{n-1}
\frac{l/n}{\left[\sin\left(\pi{l}/{n}\right) \right]^{2 }} =
\frac16 (n^2-1)\,,
\end{equation}
so that $ P_1'=1/3$ (but we remember that this number has no physical meaning, because it is 
exactly canceled by the denominator to give ${\cal F}'_1(x)=0$).

The other easily solvable case is $\alpha=1/2$, for which we have
\begin{equation}
P_n =
\sum_{l=1}^{n-1}
\frac{l/n}{ \sin\left(\pi{l}/{n}\right) }\,.
\end{equation}
We can use
\be
\int_0^\infty\frac{x^{\mu}}{(1+x)^2} dx=\frac{\pi\mu}{\sin\pi\mu}
\ee
with $\mu=l/n$, to have
\bea
P_n & =&\frac{1}\pi
\int_0^\infty\frac{dx}{(1+x)^2} \sum_{l=1}^{n-1} x^{l/n }=
\frac{1}\pi \int_0^\infty\frac{dx}{(1+x)^2}\frac{x-x^{1/n}}{x^{1/n}-1}\nonumber\\&=&
\frac{n}{\pi} \int_0^\infty (\coth x \tanh (nx)-1) dx \,,
\eea
that is the desired analytic continuation. 
From this we have
\be
 P'_1=\frac1\pi\int_0^\infty \frac{x \ln x}{(x-1)(1+x)^2}dx=
 \frac1\pi\int_0^\infty \frac{x }{\sinh x\cosh x}dx=
 \frac\pi8\,.
\ee
It is also possible to give other simple formulas for all integer values of $2\alpha$, but they 
are not values of physical interest. Let us just mention that for $\alpha=0$ we trivially have
$ P_n=(n-1)/2$ with $ P'_1=1/2$.

\begin{figure}[t]
\includegraphics[width=0.76\textwidth]{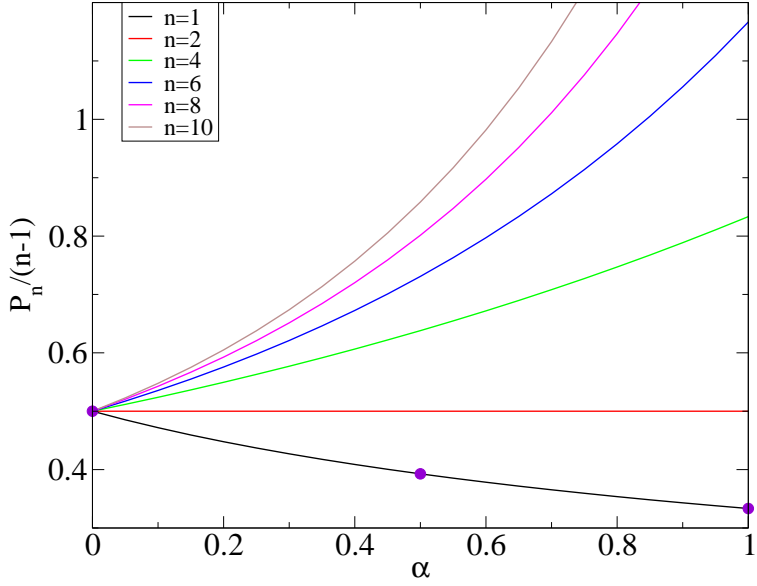}
\caption{$P_n/(n-1)$ as function of $\alpha={\rm min} (\eta,1/\eta)$ for several integral $n$ and
the analytic continuation to $n\to1$ needed for the entanglement entropy (lowest curve). 
The large points are for the three values known analytically (they perfectly agree 
with the numerical computation, confirming its correctness).}
\label{figP}
\end{figure}

For general $0<\alpha<1$, the only strategy we found is to expand the argument of the sum 
using
\be
\frac{x}{(\sin x)^\mu}=x^{1-\mu}\sum_{k=0}^\infty p_{k}x^{2k} \,,
\ee
with $p_k$ known (also to Mathematica). 
Using this expansion we have
\bea
P_n &=&\frac{1}\pi
\sum_{l=1}^{n-1}\frac{\pi l/n}{\left[ \sin\left(\pi {l}/{n}\right)\right]^{2\alpha} }=
\frac1\pi \sum_{k=0}^\infty p_k \left(\frac\pi{n}\right)^{1+2(k-\alpha)} 
\sum_{l=1}^{n-1} l^{1+2(k-\alpha)}\nonumber\\&=&
\frac1\pi 
\sum_{k=0}^\infty  p_k \left(\frac\pi{n}\right)^{1+2(k-\alpha)}  H_{n-1}^{(-2 (k-\alpha)-1)}\,,
\eea
where  the harmonic number $H_{r}^{(\mu)}$ is the natural analytic continuation of the above sum 
(defined e.g. in terms of the Riemann $\zeta$ as $H_{z}^{(\mu)}=\zeta(\mu)-\zeta(\mu,z+1)$). 
This sum is still not analytically possible, and does not provide a proper analytic continuation yet. 
In fact, it is easy to check that only for integer $n$ it gives a convergent sum (that can be 
simply  truncated at a given order to have the desired precision). 
For non-integral values, the sum is asymptotic and must be resummed. 
We explicitly show how to do this only for $ P'_1$ needed for the entanglement entropy. 
Taking the derivative wrt to $n$, we formally get
\be
 P'_1= \sum_{k=0}^\infty p_k \pi^{2(k-\alpha)}(1-2(k-\alpha))\zeta(2(\alpha-k))\,.
\ee
This sum is asymptotic. 
We then introduce the function
\be\fl
{\cal P}(z)= \sum_{k=0}^\infty p_k z^{2(k-\alpha)}(1-2(k-\alpha))\zeta(2(\alpha-k))
\equiv \sum_{k=0}^\infty P_k z^{2(k-\alpha)}\,,
\ee
with the coefficient $P_k$ growing like a factorial for large $k$. 
By definition we have $P'_1={\cal P}(\pi)$.
The $\alpha$-Borel transform is 
\be
B_{\cal P}(t)=  \sum_{k=0}^\infty \frac{P_k}{\Gamma(2(k-\alpha+1))} t^{2(k-\alpha)}\,,
\label{borel}
\ee
that provides a convergent sum.
In case we would have been able to perform this sum analytically,  the original function 
would be given by the anti-Borel transform 
\be
{\cal  P}(z)=\frac1z \int_0^\infty e^{-t/z} t^{-2\alpha}B_{\cal P}(t) dt\,,
\label{antiB}
\ee
as it can be checked by expanding in $z$. 
(We used the $\alpha$-Borel transform to cancel the effect of the singularity of the Borel 
transform in the origin, however this has no importance for the numerical results). 
The analytic sum can not be performed, and so one should find an effective approximation 
of the sum (\ref{borel}) that makes the integral (\ref{antiB}) finite. 
There are several standard methods to do this and we use the Pad\`e 
approximation of the series, that is a ratio of polynomial of order $N$ and $D$ for numerator
and denominator respectively. We consider in the integral (\ref{antiB}) large enough 
$N$ and $D$, so that the value of $ P'_1$ does not change (at the required 
accuracy) when still increasing them. 
Within this procedure we got the values of $ P'_1$ for any $0<\alpha<1$ that are reported 
in the Fig. \ref{figP} together with $P_n/(n-1)$ for integer $n$.

\end{document}